\documentclass[aoas]{imsart}

\RequirePackage{amsthm,amsmath,amsfonts,amssymb}
\RequirePackage[authoryear]{natbib}
\RequirePackage[colorlinks,citecolor=blue,urlcolor=blue]{hyperref}%% uncomment this for coloring bibliography citations and linked URLs
\RequirePackage{graphicx}%% uncomment this for including figures
\usepackage{floatrow}
\usepackage[ruled,vlined]{algorithm2e}
\usepackage{enumerate}
\usepackage{xcolor}

\startlocaldefs
\theoremstyle{plain}
\newtheorem{axiom}{Axiom}
\newtheorem{claim}[axiom]{Claim}

\newfloatcommand{capbtabbox}{table}[][\FBwidth]

\newcommand\unfootnote[1]{%
  \begingroup
  \renewcommand\thefootnote{}\footnote{#1}%
  \addtocounter{footnote}{-1}%
  \endgroup
}

\endlocaldefs

\begin{document}

\begin{frontmatter}
\title{A Bayesian Model of Underreporting for Sexual Assault on College Campuses}
%\title{A sample article title with some additional note\thanksref{T1}}
\runtitle{Bayesian Model of Underreporting for Sexual Assault} %goes in header of odd-numbered pages
%\thankstext{T1}{A sample of additional note to the title.}

\begin{aug}
%%%%%%%%%%%%%%%%%%%%%%%%%%%%%%%%%%%%%%%%%%%%%%%
%% Only one address is permitted per author. %%
%% Only division, organization and e-mail is %%
%% included in the address.                  %%
%% Additional information can be included in %%

\author[A]{\fnms{Casey}~\snm{Bradshaw}\ead[label=e1]{cb3431@columbia.edu}}%,
\and
\author[A]{\fnms{David M.}~\snm{Blei}\ead[label=e2]{david.blei@columbia.edu}}

%%%%%%%%%%%%%%%%%%%%%%%%%%%%%%%%%%%%%%%%%%%%%%
%% Addresses                                %%
%%%%%%%%%%%%%%%%%%%%%%%%%%%%%%%%%%%%%%%%%%%%%%
\address[A]{Department of Statistics, Columbia University\printead[presep={,\ }]{e1}\printead[presep={,\ }]{e2}}

%\address[B]{Department of Statistics and Department of Computer Science, Columbia University\printead[presep={,\ }]{e2}}

\end{aug}

\begin{abstract}
\par In an effort to quantify and combat sexual assault, US colleges and universities are required to disclose the number of reported sexual assaults on their campuses each year. However, many instances of sexual assault are never reported to authorities, and consequently the number of reported assaults does not fully reflect the true total number of assaults that occurred; the reported values could arise from many combinations of reporting rate and true incidence. In this paper we estimate these underlying quantities via a hierarchical Bayesian model of the reported number of assaults. We use informative priors, based on national crime statistics, to act as a tiebreaker to help distinguish between reporting rates and incidence. We outline a Hamiltonian Monte Carlo (HMC) sampling scheme for posterior inference regarding reporting rates and assault incidence at each school, and apply this method to campus sexual assault data from 2014-2019. Results suggest an increasing trend in reporting rates for the overall college population during this time. However, the extent of underreporting varies widely across schools. That variation has implications for how individual schools should interpret their reported crime statistics.
\end{abstract}


\end{frontmatter}
%%%%%%%%%%%%%%%%%%%%%%%%%%%%%%%%%%%%%%%%%%%%%%
%% Please use \tableofcontents for articles %%
%% with 50 pages and more                   %%
%%%%%%%%%%%%%%%%%%%%%%%%%%%%%%%%%%%%%%%%%%%%%%
%\tableofcontents

%%%%%%%%%%%%%%%%%%%%%%%%%%%%%%%%%%%%%%%%%%%%%%
%%%% Main text entry area:

%%%%%%%%%%%%%%%%%%%%%%%%%%%%%%%%%%%%%%%%%%%%%%
%%%%%%%%%%%%%%%%%%%%%%%%%%%%%%%%%%%%%%%%%%%%%%
\section{Introduction}
\label{sec:intro}

\par Sexual assault on college campuses is a pressing public health concern in the United States, prompting government action such as the 2014 formation of the White House Task Force to Protect Students from Sexual Assault \citep{obama2014taskforce} and the "It's on us" awareness campaign \citep{somanader2014itsonus}. Per the Clery Act of 1990, US colleges and universities are required to report annual crime statistics for their campuses and adjacent areas. These crime statistics include the number of reported sexual assaults, but because sexual assault is widely believed to be underreported, these figures do not tell the whole story. The number of reported assaults in the Clery data could arise from many different combinations of reporting rates and true underlying numbers of assaults. Without further assumptions, %we cannot distinguish between scenarios with many assaults and low reporting rates, and scenarios with few assaults but high reporting rates
it is equally plausible that a given school has a large number of assaults and a low reporting rate, or few assaults and a high reporting rate. 

\par To estimate the true incidence and reporting rates of sexual assault in the US college population, we construct a hierarchical Bayesian model of the reported data, together with a Markov Chain Monte Carlo (MCMC) sampling scheme for posterior inference. More generally, underreported count data arises in many disciplines, with applications including criminology \citep{moreno1998estimating, fernandez2019untangling} as well as epidemiology \citep{bailey2005modeling, bracher2021marginal}, traffic safety \citep{kumara2005application, ma2010bayesian}, and economics \citep{winkelmann1996markov, fader2000note}. The central challenge in such problems is to disentangle the per-school reporting rate from the true latent counts. This issue can be remedied either through the use of validation data not subject to underreporting, or by using domain knowledge to assign more informative prior distributions to latent variables. In the case of campus sexual assault, fully observed validation data is unavailable, and so we use national crime statistics as a source of external information. The absence of fully observed validation data also has implications for model assessment. Without knowledge of the true total number of assaults, we use predictive checks applied to held-out data to demonstrate the fitted model's compatibility with the data.

\par In general, we find that inference for the latent total number of assaults and reporting rate at a given school offer a clearer picture of the campus environment than the reported figures alone. In particular, these estimates can suggest the relative contributions of reporting rate and incidence to year-over-year changes in the reported number of assaults. This is a relevant line of inquiry, given that ex ante an increase in the reported number of assaults could equally be explained by an increase in the reporting rate (a positive outcome, indicating that students are aware of campus support resources) or an increase in the true total number of assaults (a bad outcome).

\par The remainder of this paper is structured as follows. Section \ref{sec:related_work} reviews related work, followed by an overview of the dataset in Section \ref{sec:data}. The proposed modeling methodology is outlined in Section \ref{sec:model}. Section \ref{sec:results} discusses modeling results and insights, and Section \ref{sec:conclusion} offers concluding remarks and directions for future work.

%%%%%%%%%%%%%%%%%%%%%%%%%%%%%%%%%%%%%%%%%%%%%%
%%%%%%%%%%%%%%%%%%%%%%%%%%%%%%%%%%%%%%%%%%%%%%

\section{Related Literature}
\label{sec:related_work}

\par There are various perspectives on underreported count data. Some studies consider underreporting as a type of censoring. In this formulation, each observed value is possibly right-censored and has an accompanying latent censoring indicator. \citet{de2017random} takes this view and assigns a covariate-dependent prior distribution to the latent censoring indicators, including the censoring indicators as a target for posterior inference.
\par Underreporting also has conceptual overlap with positive-unlabeled (PU) learning. In PU learning, data consist of labeled positive cases and unlabeled cases which may be positive or negative \citep{letouzey2000learning}. The recorded number of positive cases is thus a lower bound for the true number of positive cases, which can be viewed as an example of underreporting. The application of PU learning techniques to underreported data is explored in \citet{shanmugam2021quantifying} and \citet{wu2023collecting}.
\par Within the broader literature on underreported count data models, one particular line of work is most relevant to modeling reports of sexual assault. These modeling approaches begin with an assumption that the latent true number of events, $z$, follows a Poisson distribution with rate parameter $\lambda$. It is common to further suppose that each of these $z$-many events may or may not be reported, independently of all others, with reporting probability $p$. The number of reported events $x$, where $x \leq z$, is the data value we observe.
\par \citet{moreno1998estimating} articulates the identifiability issue in this setup: marginally, the observed data $x_{1}, \dots, x_{n}$ follow a $\textrm{Poisson}(\lambda p)$ distribution, so without further constraints or assumptions, one cannot distinguish among the possible pairs $(\lambda, p)$ with the same product. \citet{moreno1998estimating} further demonstrate that placing uninformative prior distributions on $\lambda$ and $p$ is insufficient to produce useful inference. For instance, the choice of a uniform prior on $p$ and a Jaynes \citep{jaynes1968prior} or Jeffreys \citep{jeffreys1946invariant} prior on $\lambda$ results in infinite posterior expectations for $\lambda$ and $z$ given the observed data. In some research applications this identifiability issue is remedied through the use of validation data not subject to underreporting \citep{powers2010bayesian,dvorzak2016sparse}. An alternative approach is to use external information to assign informative prior distributions, as \citet{schmertmann2018bayesian} and \citet{stoner2019hierarchical} do for the reporting probability. Because external information is available regarding both the incidence and reporting rate of sexual assault, we take this latter approach, choosing informative priors for both components of the model.

\par Earlier work on underreporting models explored the somewhat simpler case of i.i.d. observations governed by a single reporting probability and incidence parameter $\lambda$ \citep{moreno1998estimating, fader2000note}. In this paper we adopt the more recent extension of works such as \citet{dvorzak2016sparse} and \citet*{stoner2019hierarchical}, allowing $p$ and $\lambda$ to vary across units. When modeling $\lambda_{i}$ and $p_{i}$, we introduce school features as covariates to account for demographic differences among schools, as detailed in Section \ref{sec:model}.

\par Posterior distributions for latent variables in this type of underreporting model are generally intractable. However, choosing conditionally conjugate priors for $\lambda$ and $p$ (Gamma and Beta distributions, respectively) leads to analytic expressions for their marginal posterior distributions \citep{moreno1998estimating,fader2000note}. Alternatively, MCMC approaches to posterior inference allow greater flexibility in choice of priors in exchange for a greater computational burden. \citet{dvorzak2016sparse} use a bespoke implementation of a Gibbs-style sampler, while \citet*{stoner2019hierarchical} use slice sampling. Our paper extends this line of work by demonstrating the viability of gradient-based samplers such as HMC for posterior inference in this model setting.
\par With respect to the application context, the most relevant work comes from \citet*{fernandez2019untangling}, who model underreporting of domestic violence complaints across districts in Spain. Unlike the campus sexual assault data, observations in \citet*{fernandez2019untangling}'s dataset are dense in the time domain (quarterly for 10 years), motivating the authors to adopt time series methodology.

%%%%%%%%%%%%%%%%%%%%%%%%%%%%%%%%%%%%%%%%%%%%%%
%%%%%%%%%%%%%%%%%%%%%%%%%%%%%%%%%%%%%%%%%%%%%%

\section{Dataset}
\label{sec:data}

The dataset contains 11,369 records for 1,973 US colleges and universities over the 2014-2019 time frame \footnote{Data from 2020 and 2021 are excluded. Because the COVID-19 pandemic significantly disrupted patterns of social interaction, campus sexual assault data from this time are not well described by the same model used for other years.}, where a ``record" consists of the total number of assaults reported at a particular school in one particular calendar year. Schools located in the 50 US states, the District of Columbia, and Puerto Rico are included in this analysis. Annual campus crime statistics are furnished by the US Department of Education's Office of Postsecondary Education. All schools that receive federal financial aid funding under Title IV of the Higher Education Act of 1965 are required to submit annual campus crime reports. Most schools have records for all six years in this time frame, though 8\% appear in fewer years due to schools opening/closing or beginning/ ceasing to participate in Title IV.
\par School characteristics and demographic data are available through the National Center for Education Statistics (NCES), which publishes comprehensive institutional profile information through its IPEDS database \footnote{ \url{https://nces.ed.gov/ipeds/}}. Preprocessed data is included in the Supplementary Material; raw data is available publicly and from the corresponding author upon reasonable request. Additional dataset details are included in Appendix \ref{app:data_description}.
\par We exclude institutions with no degree-granting programs and institutions with no programs classified as "academic" in nature. We also exclude institutions with no residential housing facilities. For such schools, the school does not represent the students' community, and we would not reasonably expect students to report sexual assaults to the school.
\par
\begin{figure}[H]
\includegraphics[width=7cm]{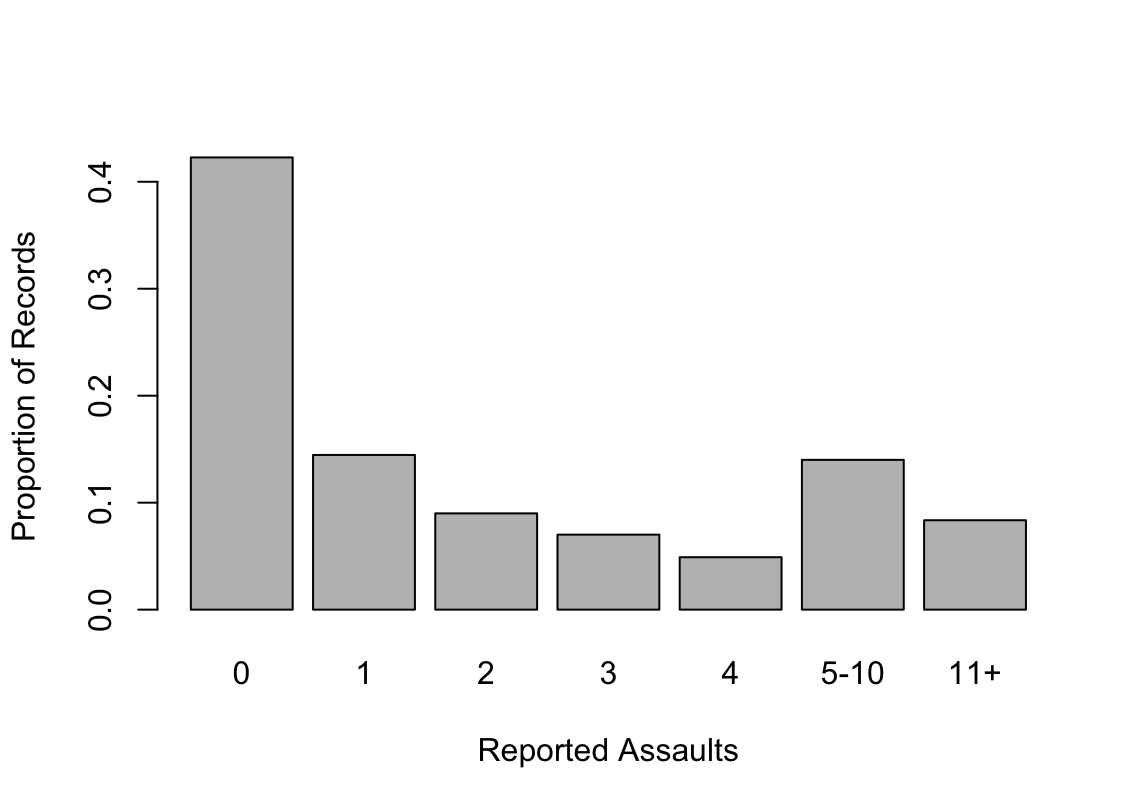}
\caption{School-level reported sexual assault data is sparse, with 42\% of records showing zero assaults, and a median of one assault.}
\label{fig:rape_barplot}
\end{figure}

 Institutions in this dataset range in size from 5 to 79,500 in-person students, with a median of 2,350. Public institutions comprise 41\% of the dataset, religiously affiliated institutions 36\%, and private, non-religiously affiliated institutions 22\%. The reported sexual assault data is notably sparse: 42\% of records show zero reported assaults, and 22\% of institutions show no assaults across all six years. The distribution of reported values, shown in Figure \ref{fig:rape_barplot}, has a long right tail, with a handful of schools reporting more than 100 assaults in a given year. This sparsity pattern is not explained by student population size alone: schools with as many as 36,000 students had zero reported assaults, while assaults \emph{were} reported at schools with as few as several hundred students.

\par
\begin{figure}[H]
\includegraphics[width=10cm]{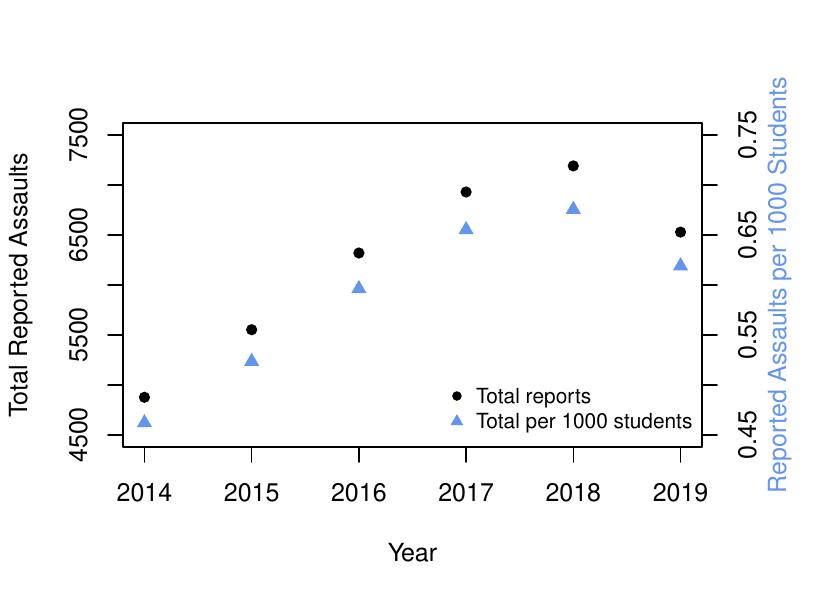}
\caption{The total number of reported assaults grew from 2014 to 2018. Total student population fluctuated relatively little, resulting in per capita reporting trends moving with the overall total number of reports.}
\label{fig:yearly_reports}
\end{figure}

 The total number of assaults reported across schools increased steadily from 2014 to 2018, growing by nearly 50\% before falling in 2019 (see Figure \ref{fig:yearly_reports}). The total in-person student population was relatively stable over this time period, remaining close to 10.6 million.

Campus crime numbers in this dataset are a direct tally of the individual assaults for which a student voluntarily made a formal report to campus authorities. Note that those reports do not arise from a survey process or broad inquiry about student experiences; on its own, this dataset contains no information about underreporting. For complementary information about reporting rates, we consider the National Crime Victimization Survey (NCVS). The NCVS is a large-scale annual survey conducted on behalf of the behalf of the US Bureau of Justice Statistics to determine incidence rates of personal and property crimes, including sexual assault. This survey dates back 50 years, and one of its primary objectives is to quantify the extent of crime not reported to authorities. By attempting to measure crimes not reported to authorities, the NCVS provides additional context which is absent from the college campus Clery Report data.

%%%%%%%%%%%%%%%%%%%%%%%%%%%%%%%%%%%%%%%%%%%%%%
%%%%%%%%%%%%%%%%%%%%%%%%%%%%%%%%%%%%%%%%%%%%%%

\section{Methodology}
\label{sec:model}

For a particular school $i$ in a particular year $j$, suppose that the true, unknown number of assaults $z_{ij}$ comes from a Poisson distribution with rate parameter $\lambda_{ij}$. Further suppose that each of those $z_{ij}$ assaults is independently reported, or not reported, with probability of reporting equal to $p_{ij}$. This produces the reported value $x_{ij}$ that we observe, according to a binomial thinning process:
\begin{equation}
    x_{ij}|z_{ij}, p_{ij} \sim \textrm{Binom}(z_{ij}, p_{ij}).\label{eq:x_given_zp}
\end{equation}

\begin{figure}[H]
\includegraphics[width=12cm]{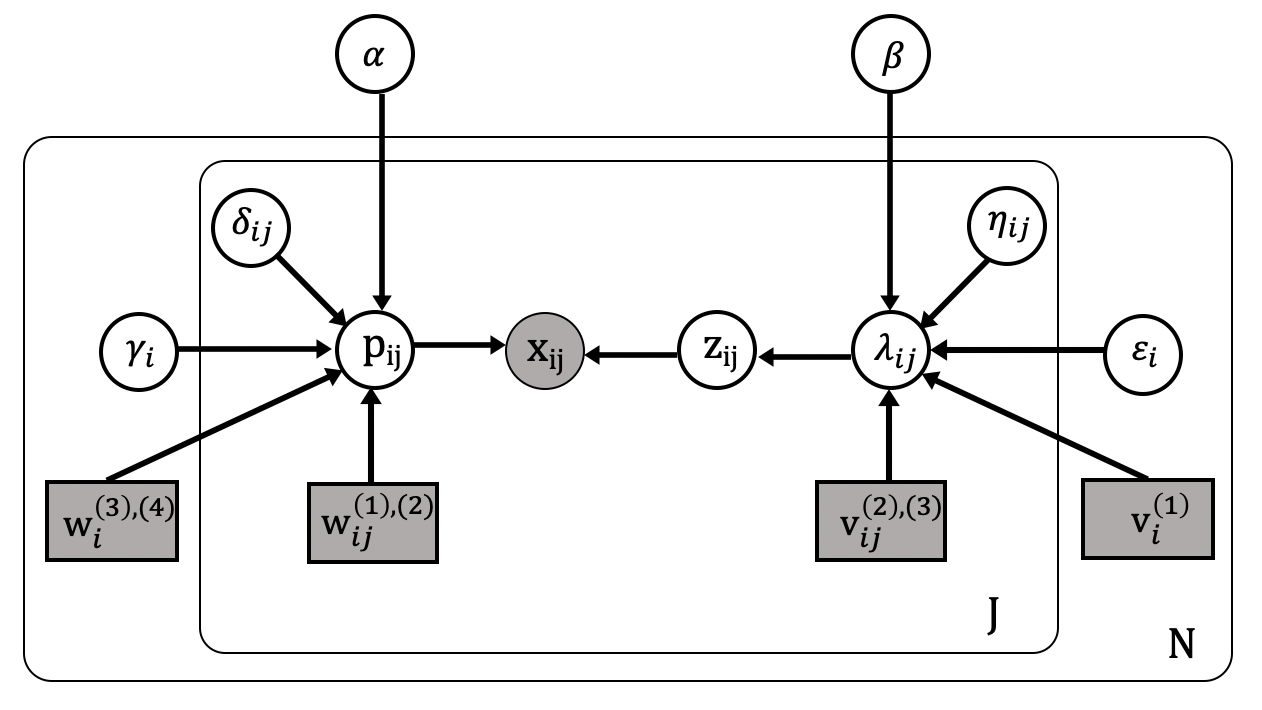}
\caption{Graphical model diagram}
\label{fig:model_diagram}
\end{figure}

With this likelihood model, we construct a hierarchical generative model for the reported campus sexual assault data. The proposed hierarchical model, diagrammed in Figure \ref{fig:model_diagram}, entails a component describing the reporting probability, and a component describing the Poisson rate governing the true number of assaults. A common approach in the underreporting literature is to model $\textrm{log}(\lambda_{ij})$ and/or $\textrm{log} \Big( \frac{p_{ij}}{1-p_{ij}} \Big)$ as deterministically equal to a linear function of covariates \citep{dvorzak2016sparse, stoner2019hierarchical, de2021estimating}. To provide additional flexibility, we instead model the Poisson rate and reporting probability as random functions of their respective covariates. Full details of each model component are discussed below.

\subsection{Modeling z}

For each record in the dataset, we suppose the true number of assaults arises from a Poisson distribution parameterized by $\lambda_{ij}$, where $\lambda_{ij}$ depends on observed characteristics \textbf{v} of school $i$ in year $j$ as follows:

\begin{eqnarray}
z_{ij}|\lambda_{ij} &\sim& \textrm{Poisson}(\lambda_{ij}) \label{eq:z}\\
\log(\lambda_{ij}) &=& \beta_{0, \textrm{v}_{i}^{(1)}} + \beta_{1} \textrm{v}_{ij}^{(2)} + \beta_{2} \textrm{v}_{ij}^{(3)} + \varepsilon_{i} + 
\eta_{ij} \label{eq:lambda}\\
\beta_{0,1}, \beta_{0,2}, \beta_{0,3}  &\sim& N(-5.5, 0.5) \nonumber\\
\beta_{1} &\sim& N(1,0.1) \nonumber\\
\beta_{2} &\sim& N(0,4) \nonumber\\
\varepsilon_{i} &\sim& N(0, 0.75) \nonumber\\
\eta_{ij} &\sim& N(0, 0.1) \nonumber
\end{eqnarray}

The covariates associated with $\lambda$ are:
\begin{center}
\begin{tabular}{ c l }  
 $\textrm{v}_{i}^{(1)}$:  & degree of urbanization of school $i$'s campus  \\
  & \hspace{2cm} 1 = urban  \\
  & \hspace{2cm} 2 = suburban\\
  & \hspace{2cm} 3 = rural\\
 $\textrm{v}_{ij}^{(2)}$:  & log(number of in-person students) for school $i$ in year $j$  \\
  $\textrm{v}_{ij}^{(3)}$: & (fraction of women in student body at school $i$ in year $j$ - 0.5)$^2$  \\
\end{tabular}
\end{center}

\par This generative model considers in-person student population, degree of campus urbanization, and student body gender composition\footnote{The gender categories reported by NCES are currently "male" and "female".} as covariates of interest. The functional form of equations \eqref{eq:z} - \eqref{eq:lambda} implies that for the expected number of assaults we have:
\begin{eqnarray}
\mathbb{E}(z_{ij} | \beta, \mathbf{v}) & \propto & \textrm{exp}(\beta_{0, \textrm{v}_{i}^{(1)}} + \beta_{2} \textrm{v}_{ij}^{(3)}) \textrm{exp}(\beta_{1} \log(\textrm{student population}_{ij})) \nonumber\\
 & \propto&  (\textrm{student population}_{ij})^{\beta_1}\label{eq:powerlaw}.
\end{eqnarray}

 This modeling choice allows for the possibility that the number of assaults does not scale linearly with the number of in-person students. Because the number of interpersonal interactions does not always scale linearly with community size, this type of "power law" dynamic has been observed for various social phenomena, including crime \citep{bettencourt2007growth, chang2018population}.
\par Incorporating separate intercepts in \eqref{eq:lambda} allows for different incidence patterns at urban, suburban, and rural schools. This decision is motivated by the national differences in violent crime victimization rates across locales of varying urbanization \citep*{ncvs13, ncvs14, ncvs15, ncvs16, ncvs17, ncvs19}. 
\par Gender composition of the student body is included as a covariate because most perpetrators of sexual assault are men, while most victims are women \citep{sinozich2014rape}. We center the proportion of women in the student body by subtracting 0.5, and square this value, to allow for the possibility that schools with more extreme gender imbalances may have different rates of assault.

\par
\begin{figure}[H]
\begin{floatrow}
\capbtabbox{%
  \begin{tabular}{ c  c  c  c  } 
\hline
  & 18-20 & 21-24 & 12+  \\
\hline
2014 \vline & 1.2* & 3.8 & 1.1  \\
2015 \vline & 3.2 & 1.6* & 1.6  \\
2016 \vline & 3.8 & 1.8 & 1.1  \\
2017 \vline & 8.6 & 2.4 & 1.4  \\
2018 \vline & 10.1 & 6.0 & 2.7 \\
2019 \vline & 7.1 & 6.4 & 1.7  \\
\hline
 & & & \\
 & & & \\
 & & & \\
\end{tabular}
\label{table:incidence}
}{%
  \caption{Sexual Assault Incidence per 1000 persons in the US, by age group}%
}
\ffigbox{%
  \includegraphics[width=7cm]{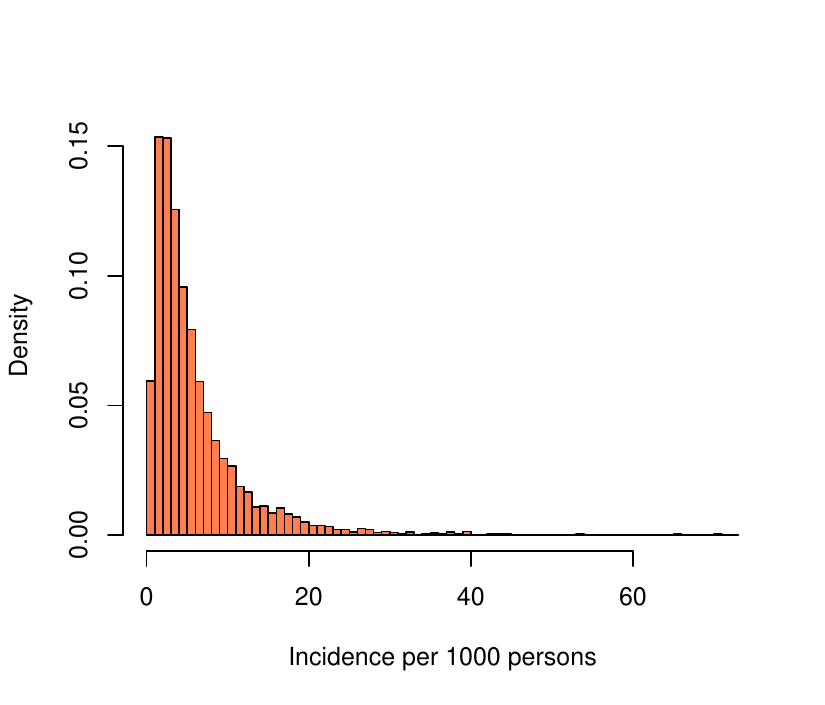}%
}{%
  \caption{Prior distributions on $\beta_{0}$, $\varepsilon$, and $\eta$ imply a distribution over incidence centered at 4.1 per 1000 persons, with a central 50\% range of 2.3 to 7.4.}%
\label{fig:prior_incidence}
}
\end{floatrow}
\end{figure}

 Per-school intercepts $\varepsilon_{i}$ allow for between-school differences in assault incidence patterns which are not otherwise captured by covariates. Per-record noise terms $\eta_{ij}$ allow for within-school differences across years. Supposing that $\beta_{1}=1$ (linear relationship between student population and number of assaults) and $\beta_{2}=0$ (no relationship between student body gender composition and expected number of assaults), priors on $\varepsilon$, $\eta$, and intercepts $\beta_{0}$ are set so that the resulting distribution over assaults per 1000 persons roughly corresponds with external estimates from the National Crime Victimization Survey (NCVS). The resulting distribution is displayed in Figure \ref{fig:prior_incidence}. Note that, conditional on covariates, the Poisson rate $\lambda_{ij}$ follows a lognormal distribution, leading to a number of assaults $z_{ij}$ whose marginal distribution is overdispersed relative to a Poisson.

\unfootnote{$^{*}$ Per BJS: "[estimate is] based on 10 or fewer sample cases, or coefficient of variation is greater than 50\%"}

The NCVS is a large sample survey conducted annually on behalf of the US Bureau of Justice Statistics to determine incidence rates of personal and property crimes. This survey includes crimes reported and not reported the police. Over the 2014-2019 time period, the estimated incidence of sexual assault in the overall population aged 12 and older ranged from 1.1 to 2.7 per 1000 persons. As shown in Table \ref{table:incidence}, estimated incidence skewed higher in the 18-20 and 21-24 age groups, which are particularly relevant to the college student population.

\subsection{Modeling p}
As formulated in Equation \ref{eq:x_given_zp}, the reported number of assaults arises from the latent true number of assaults $z$ and the reporting probability $p$. Having discussed a model for $z$, we now turn to modeling the reporting probability $p$. Of the total $z_{ij}$ assaults occurring at school $i$ in year $j$, suppose that each individual assault is reported, independently, with probability $p_{ij}$, where $p_{ij}$ depends on observed characteristics \textbf{w} of school $i$ in year $j$ as follows:
\begin{eqnarray}
\textrm{log} \Big( \frac{p_{ij}}{1-p_{ij}} \Big) &=& \alpha_{0}+\alpha_1 \textrm{w}_{i}^{(1)} + \alpha_{2} \textrm{w}_{i}^{(2)}+ \alpha_{3} \textrm{w}_{ij}^{(3)} + \alpha_{4} \textrm{w}_{ij}^{(4)}+ \gamma_{i} + \delta_{ij} \label{eq:p}\\
\alpha_{0} &\sim& N(-1.25, 0.5)\nonumber\\
\alpha_{1}, \alpha_{2}, &\sim& N(0,2)\nonumber\\
\alpha_{3}, \alpha_{4} &\sim& N(0,4)\nonumber\\
\gamma_{i} &\sim& N(0, 1.25)\nonumber\\
\delta_{ij} &\sim& N(0, 0.5).\nonumber
\end{eqnarray}

The covariates associated with $p$ are:
\begin{center}
\begin{tabular}{ c l }  
 $\textrm{w}_{i}^{(1)}$: & school $i$ issues associate degrees only ($\in \{ 0,1\}$)\\
 $\textrm{w}_{i}^{(2)}$: & school $i$ substantially engaged in religious instruction ($\in \{ 0,1\}$)\\
 $\textrm{w}_{ij}^{(3)}$:  & (women as fraction of student body at school $i$ in year $j$ - 0.5)  \\
 $\textrm{w}_{ij}^{(4)}$:  & Pell grant recipients as fraction of student body at school $i$ in year $j$, median centered
\end{tabular}
\end{center}
\vspace{12pt}

\begin{figure}[H]
\includegraphics[width=6cm]{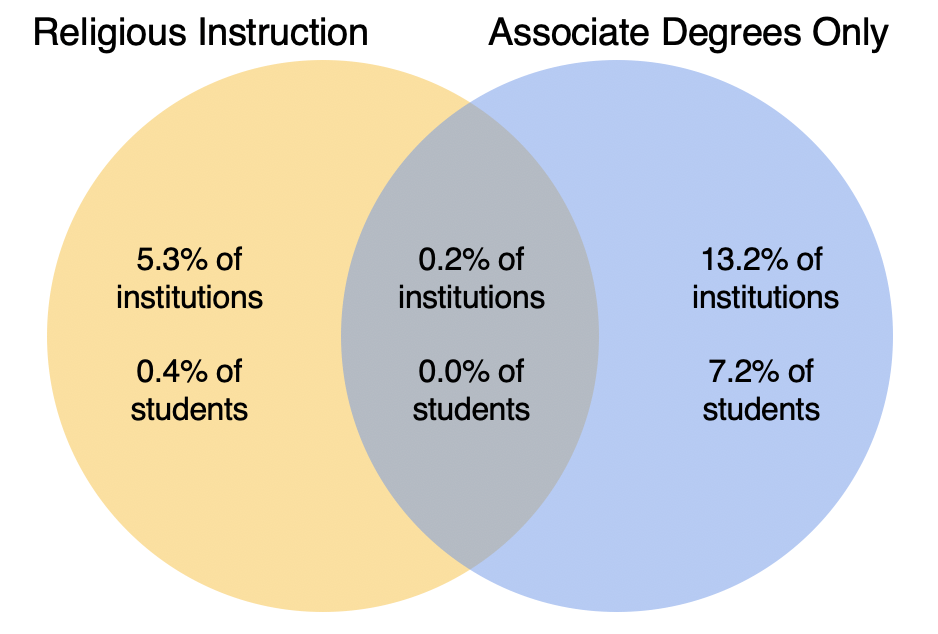}
\caption{Junior colleges and institutions of religious instruction comprise a minority of the dataset}
\label{fig:venn}
\end{figure}

The reporting probability at school $i$ in year $j$ accounts for two school-level covariates that are fixed across years, and two that vary. The first fixed school-level covariate is $\textrm{w}_{i}^{(1)}$, an indicator for whether school $i$ is a junior college, issuing only associate degrees as opposed to bachelor degrees or higher. The reporting probability also incorporates an indicator $w_{i}^{(2)}$ for whether school $i$ is substantially engaged in religious instruction. A theological seminary, for instance, would satisfy this definition, whereas a religiously affiliated liberal arts college would not. The inclusion of these school-level random effects encourages the probabilities for an individual school to exhibit similarity across years. Junior colleges comprise roughly 13\% of institutions in the dataset, and account for 7\% of total in-person students. Institutions of religious instruction make up a relatively smaller portion of the dataset, and generally have small student populations. These two categories of institution have negligible overlap, as detailed in Figure \ref{fig:venn}.

Gender composition of a school's student body also contributes to the reporting probability, and varies across years. Many institutions in this dataset have approximately gender-balanced student populations. Across records, the proportion of female students is centered around 57\%, with smaller peaks at each end of the spectrum corresponding to all-male and all-female schools. The final covariate incorporated in the reporting rate is the fraction of Pell grant recipients in the student body. Pell grants are federal financial aid awarded to students on the basis of exceptional financial need. This variable is included as a proxy for socioeconomic status, which research indicates may be correlated with reporting rates \citep{fisher2003reporting, sabina2014campus}. The median percentage of undergraduates receiving Pell grants is 36\%, and values are more widely spread over the unit interval compared to the gender composition composition covariate.

\par
\begin{figure}[H]
\begin{floatrow}
\capbtabbox{%
  \begin{tabular}{ c  c  c  c  }  
\hline
 & 18-20 & 21-24 & 12+ \\
\hline
2014 \vline & 0.0* & 16.48 & 33.6  \\
2015 \vline & 10.6* & 26.0* & 32.5  \\
2016 \vline & 10.2* & 31.0* & 23.2  \\
2017 \vline & 45.3 & 38.3* & 40.4  \\
2018 \vline & 11.0* & 10.2* & 24.9 \\
2019 \vline & 17.5* & 24.7* & 33.9  \\
\hline
 & & & \\
 & & & \\
 & & & \\
\end{tabular}
\label{table:reporting}
}{%
  \caption{Percentage of Sexual Assaults Reported to Police, by age group}%
}
\ffigbox{%
  \includegraphics[width=7cm]{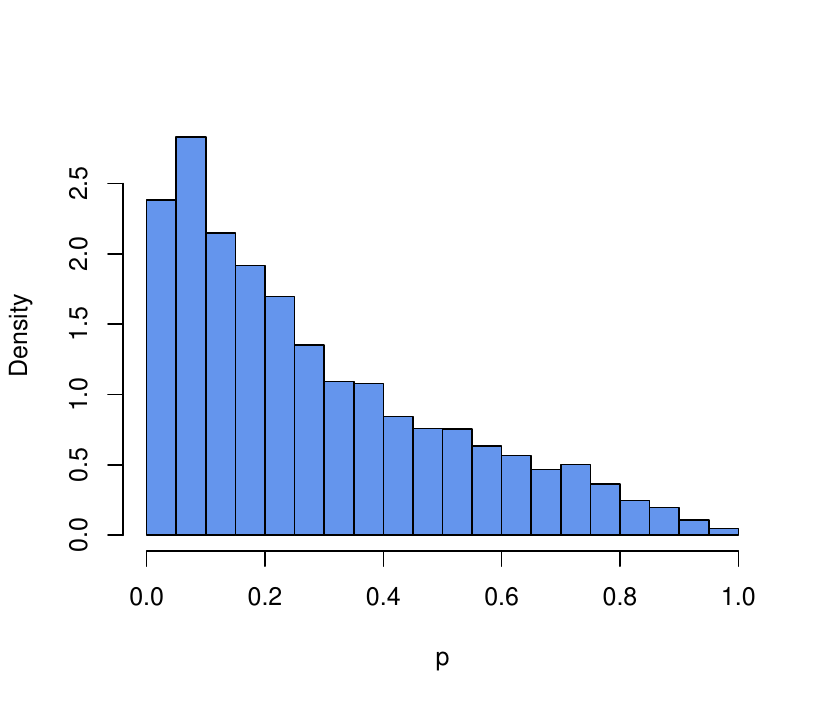}%
}{%
  \caption{Prior distributions on $\alpha_{0}$, $\gamma$, and $\delta$ imply a distribution over reporting probability centered at 22\%, with a central 50\% range of 10\% to 43\%.}%
\label{fig:prior_ps}
}
\end{floatrow}
\end{figure}

 As in the model of Poisson rate $\lambda_{ij}$, per-school intercepts $\gamma_{i}$ in Equation \ref{eq:p} allow for between-school differences in assault incidence patterns not captured by covariates, while per-record noise terms $\delta_{ij}$ allow for within-school differences across years. Supposing that $\alpha_1=\alpha_2=\alpha_3=\alpha_4=0$, priors on $\varepsilon$, $\eta$, and intercepts $\beta_{0}$ are chosen to be broadly consistent with NCVS estimates of the percentage of sexual assaults reported to police. Over the 2014-2019 time period, the estimated rate of reporting to police ranged from 23\% to 40\% in the overall 12+ population, while reporting rates for the 18-20 and 21-24 age groups tended to be lower (see Table \ref{table:reporting}). The induced prior distribution over values of $p$, when $\alpha_1=\alpha_2=\alpha_3=\alpha_4=0$, is depicted in Figure \ref{fig:prior_ps} below.

%Furthermore, earlier research indicating that roughly 76\% of violent crimes occurring at schools were not reported to police, and that sexual assault had a relatively lower rate of reporting than other violent crimes \citep*{notreportedtopolice}

In these models of $z$ and $p$, global coefficients $\alpha$ and $\beta$ encourage the sharing of information across schools, while school-level terms $\mathbf{\gamma}$ and $\mathbf{\varepsilon}$ capture local properties of individual schools. This partial pooling model structure is a compromise between complete pooling, in which no school-level differences are permitted ($\gamma$ and $\varepsilon$ omitted), and no pooling, in which each school $i$ is modeled separately from all others, with its own coefficient vectors $\alpha_i$ and $\beta_i$. Compared to the complete-pooling and no-pooling alternatives, the partial pooling model assigns higher likelihood to held-out observational data- refer to Appendix \ref{app:pooling} for details.

%%%%

%
\subsection{Inference}

The model does not admit closed form expressions for posterior quantities of interest. We draw samples from the joint posterior distribution over latent variables via MCMC. The true number of assaults $z_{ij}$ is a discrete latent variable, which presents an obstacle to gradient-based sampling schemes such as Hamiltonian Monte Carlo (HMC). However, this can be circumvented by marginalizing out $z_{ij}$. Instead of working with $z_{ij}$, one can take advantage of the fact that conditional on the event rate $\lambda_{ij}$ and reporting probability $p_{ij}$, but not the true number of assaults $z_{ij}$, the reported number of assaults follows a Poisson distribution:
$$x_{ij} | \lambda_{ij}, p_{ij}, \mathbf{w}_{ij}, \mathbf{v}_{ij} \sim \textrm{Poisson}(\lambda_{ij} p_{ij}).$$
See Appendix \ref{app:derivations} for derivation. This marginalizes out the discrete latent variable $z_{ij}$, allowing for gradient-based MCMC sampling from the posterior $p(\boldsymbol\lambda, \mathbf{p} | \mathbf{x}, \mathbf{w}, \mathbf{v})$. However, the latent true number of assaults $z_{ij}$ is also a quantity of interest. To address this, first note the conditional distribution of the number of unreported assaults, $u_{ij}$ (where $u_{ij} \overset{\underset{\mathrm{def}}{}}{=} z_{ij} - x_{ij}$): 
$$u_{ij} | x_{ij}, \lambda_{ij}, p_{ij} \sim \textrm{Poisson}(\lambda_{ij}(1-p_{ij})).$$

Because $u_{ij}$ is conditionally independent of the covariates $\mathbf{v}_{ij}$ and $\mathbf{w}_{ij}$ given latent parameters $\lambda_{ij}$ and $p_{ij}$, it is equivalently true that 
$$u_{ij} | x_{ij}, \lambda_{ij}, p_{ij}, \mathbf{w}_{ij}, \mathbf{v}_{ij} \sim \textrm{Poisson}(\lambda_{ij}(1-p_{ij})).$$
Each MCMC sample $(\lambda_{ij}^{(s)}, p_{ij}^{(s)})$ drawn from (approximately) $p(\lambda_{ij}, p_{ij} | x_{ij}, \mathbf{w}_{ij}, \mathbf{v}_{ij})$ can be augmented by sampling a corresponding $u_{ij}^{(s)}$ from $p(u | x_{ij}, \lambda_{ij}, p_{ij})$. The sampled value $u_{ij}^{(s)}$ can then be added to the observed count $x_{ij}$ to produce a sample for the latent count $z_{ij}^{(s)}$. Altogether, the sample $(\boldsymbol\lambda^{(s)}, \mathbf{p}^{(s)}, \mathbf{z}^{(s)})$ constitutes an MCMC sample from $p(\boldsymbol\lambda, \mathbf{p}, \mathbf{z} |\mathbf{x})$. This sampling procedure is outlined in Algorithm \ref{algo:posterior}, and can be conveniently implemented in a probabilistic programming language.  Posterior inference for this work was conducted using HMC in Stan \citep*{stanpackage}.

\begin{algorithm}[H]
\label{algo:posterior}
\SetAlgoLined
\KwIn{ \begin{itemize}
\item Data $\mathbf{x}$ \\
\item Covariates $\mathbf{w}, \mathbf{v}$
\item Index set $\mathcal{I}$ (pairs (i,j) for which the dataset contains a record $x_{ij}$  \\ 
\item MCMC algorithm \texttt{SAMPLER} (e.g. HMC)  \\
\item Markov chain length $S$
\end{itemize} }
\KwOut{ samples from the posterior $p(\mathbf{z},\boldsymbol\lambda, \mathbf{p} | \mathbf{x}, \mathbf{w}, \mathbf{v})$; }
 
 \For{s in 1:S}{
  Draw $(\boldsymbol\lambda^{(s)}, \mathbf{p}^{(s)})$ from \texttt{SAMPLER} approximating $p(\boldsymbol\lambda, \mathbf{p} 
  | \mathbf{x}, \mathbf{w}, \mathbf{v})$ \;
  \For{$(i,j)$ in $\mathcal{I}$}{
    Draw $u_{ij} \sim \textrm{Poisson}((1-p_{ij}^{(s)}) \lambda_{ij}^{(s)}) $ \;
    Set $z_{ij} = u_{ij} + x_{ij}$ \;
  }

  }
 \KwRet{$ \left\{ (z^{(s)}, \lambda^{(s)}, p^{(s)}) \right\}_{s=1}^{S}$}
 \caption{Posterior Sampling Scheme}
\end{algorithm}

\subsection{Model Assessment}

Fully-observed data including the true total number of assaults is not available, and thus cannot be used for model validation. Instead, the predictive distribution provides information about the quality of the model fit. The posterior distribution of the latent variables implies a predictive distribution over future observations $p(x^{new} | x)$ and, accordingly, over summary statistics of future observations. This motivates the notion of posterior predictive checks \citep{guttman1967use, box1980sampling, rubin1984bayesianly, gelman1996posterior}: that a satisfactory model will yield predictive distributions over key summary statistics under which the actual values observed in the dataset are not extreme. As detailed in \citet{moran2019population} and \citet{li2022calibrated}, such posterior predictive checks can be overly optimistic about model fit due to double use of the data (both to fit the model and to evaluate it). The aforementioned works propose randomly reserving a portion of the original dataset to use exclusively for model assessment, an approach that demonstrably achieves a more accurate gauge of model performance by avoiding the pitfall of data reuse. In the spirit of this suggestion, roughly 20\% of entries in the campus sexual assault dataset are held out, with the remaining 80\% used for posterior inference. Held out data points are sampled at the level of (school, year) pairs. The held out data set includes records from all years in scope (2014-2019), and does not contain data from any additional schools not seen in the main sample. Predictive samples for the held out data points are generated as follows.

\begin{algorithm}[H]
\label{algo:poppc}
\SetAlgoLined
\KwIn{ \begin{itemize}
\item Index set $\mathcal{I}$ (pairs (i,j) for which the held out dataset contains a record $x^{new}_{ij}$)  \\
\item Covariates $\mathbf{w}^{new}, \mathbf{v}^{new}$ for held out dataset \\ 
\item Posterior distribution $p(\boldsymbol\alpha, \boldsymbol\beta, \boldsymbol\gamma, \boldsymbol\varepsilon | \mathbf{x}, \mathbf{w}, \mathbf{v})$ 
\end{itemize} }
\KwOut{ Sample dataset from the predictive distribution $p( \mathbf{x}^{new} | \mathbf{x}, \mathbf{w}, \mathbf{v}, \mathbf{w}^{new}, \mathbf{v}^{new})$;}
     \BlankLine
     \BlankLine
    Draw $(\boldsymbol\alpha ^{s}, \boldsymbol\beta ^{s}, \boldsymbol\gamma ^{s}, \boldsymbol\varepsilon ^{s})$ \;
  \For{$(i,j)$ in $\mathcal{I}$}{
    Draw $\delta_{ij} \sim N(0,0.5)$ \;
    Draw $\eta_{ij} \sim N(0,0.2)$ \;
    \eIf{school $i$ $\in$ training data}{
        $\gamma_{i} \gets \gamma_{i}^{s}$  \;
        $\varepsilon_{i} \gets \varepsilon_{i}^{s}$ \; 
    }{
    Draw $\gamma_{i} \sim N(0,1)$ \;
    Draw $\varepsilon_{i} \sim N(0,0.5)$ \;
    }
    $p_{ij} \gets$ invlogit$(\alpha_{0}^{s} + \langle \boldsymbol{\alpha}^{s}, \mathbf{w_{ij}}^{new} \rangle + \gamma_{i} + \delta_{ij})$ \;
    $\lambda_{ij} \gets$ exp$(\beta_{0,v_i}^{s} +  \langle \boldsymbol{\beta}^{s}, \mathrm{v}_{ij}^{new} \rangle + \varepsilon_{i} + \eta_{ij}) $ \;
    Draw $z_{ij} \sim \textrm{Poisson}(\lambda_{ij})$ \;
    Draw $x_{ij}^{new} \sim \textrm{Binom}(z_{ij}, p_{ij})$\;
  }
  
 \KwRet{$ \mathbf{x}^{new}$}
 \caption{Predictive Distribution Sampling}
\end{algorithm}
\vspace{12pt}

As noted in Section \ref{sec:data}, each year many schools had zero reported assaults, and the median number of reports was 1. Generating 10,000 resampled datasets according to Algorithm \ref{algo:poppc} and calculating the proportion of records with zero reported assaults produces the distribution shown below in Figure \ref{fig:posterior_predictive_checks}(a). Within the held out dataset 42.1\% of records had zero reported assaults, while values ranged from roughly 38\% to 44\% in datasets sampled from the predictive distribution. The distribution of reported assault numbers has been pulled slightly away from the extreme end, such that the observed proportion of zeroes falls toward the high end of the predictive distribution. However, the median number of assaults was equal to 1 in all datasets sampled from the predictive distribution, and the true proportion of held-out records $x_{ij}$ with $\leq 1$ assaults sits comfortably within its respective predictive distribution (Figure \ref{fig:posterior_predictive_checks}(b)). Overall, the model adequately captures the bottom-heaviness of the distribution of reported assault numbers.

\begin{figure}[H]
\begin{center}
\begin{tabular}{c c}
    \includegraphics[width=6.5cm]{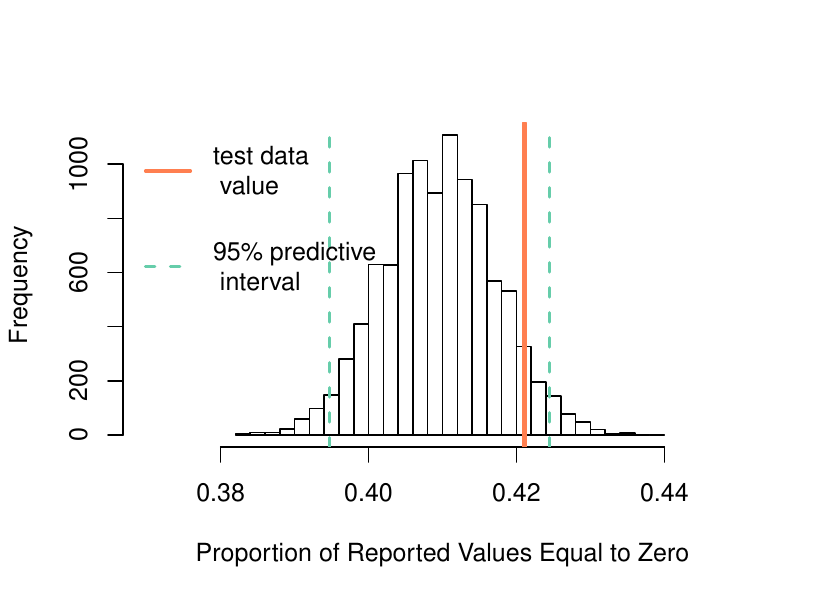} & \includegraphics[width=6.5cm]{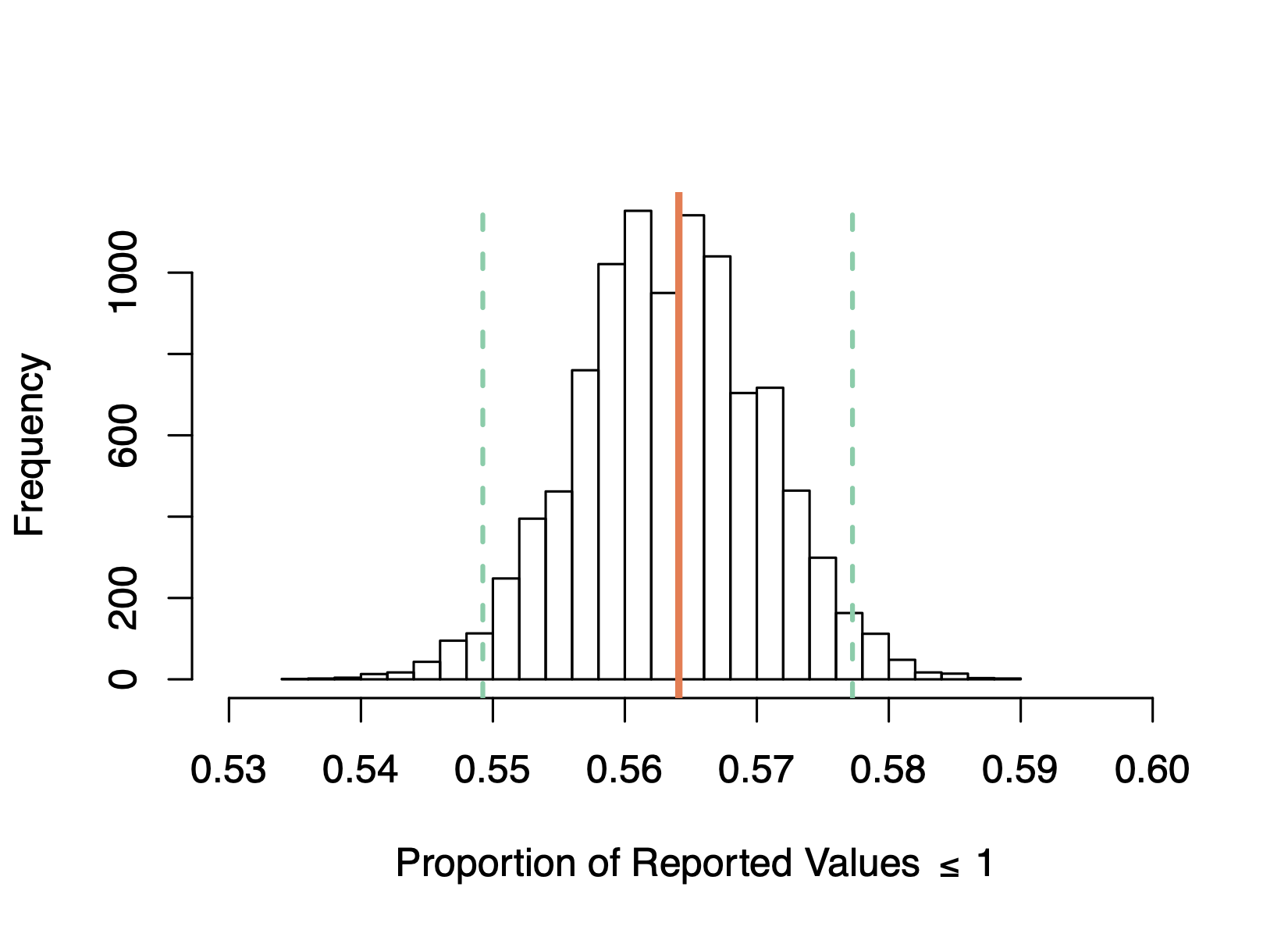} \\
    (a) & (b)
\end{tabular}
\begin{tabular}{c c}
\includegraphics[width=6.5cm]{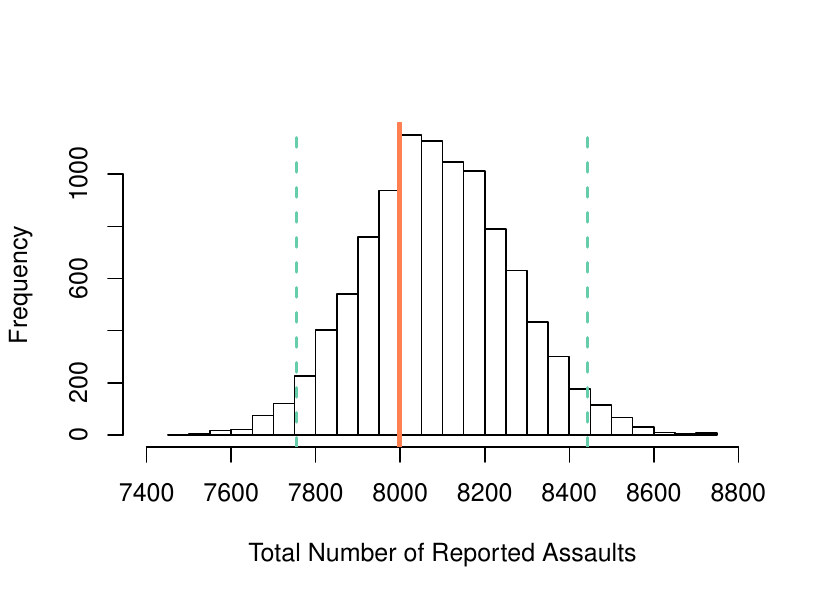} & \includegraphics[width=6.5cm]{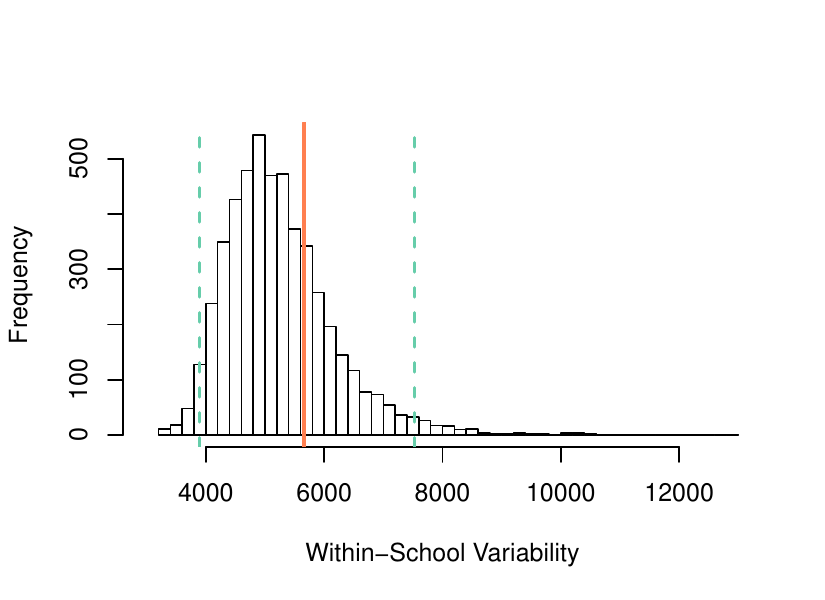}\\
(c) & (d)
\end{tabular}
    \caption{Summary statistics of the held out data are plausible under the predictive distribution. Solid lines indicate the value calculated on the held out data, dashed lines indicate the 2.5th and 97.5th percentiles of the predictive distribution; (a) Proportion of reports indicating zero assaults (b) Proportion of reports indicating $\leq 1$ assaults (c) Total number of reported assaults across all schools and all years (d) Within-school variability of reported numbers over time}
\label{fig:posterior_predictive_checks}
\end{center}
\end{figure}

\par Other quantities of interest are the total number of reported assaults $\sum_{ij} x_{ij}^{new}$ and the amount of within-school variability. The predicted number of assaults reported by an individual school should not be too rigid or too flexible over time, relative to the trends seen in the actual data. This summary statistic is quantified as $\sum_{i} \textrm{var}(x_{i \cdot}^{new})$. The total number of reported assaults and the within-school variability observed in the held out dataset are both plausible under their respective predictive distributions, shown in Figure \ref{fig:posterior_predictive_checks}(c)-(d), suggesting that the model adequately captures these characteristics of the data.

\subsection{Capabilities and Limitations}

\begin{figure}[H]
\begin{center}
\begin{tabular}{c c}
    \includegraphics[width=6.5cm]{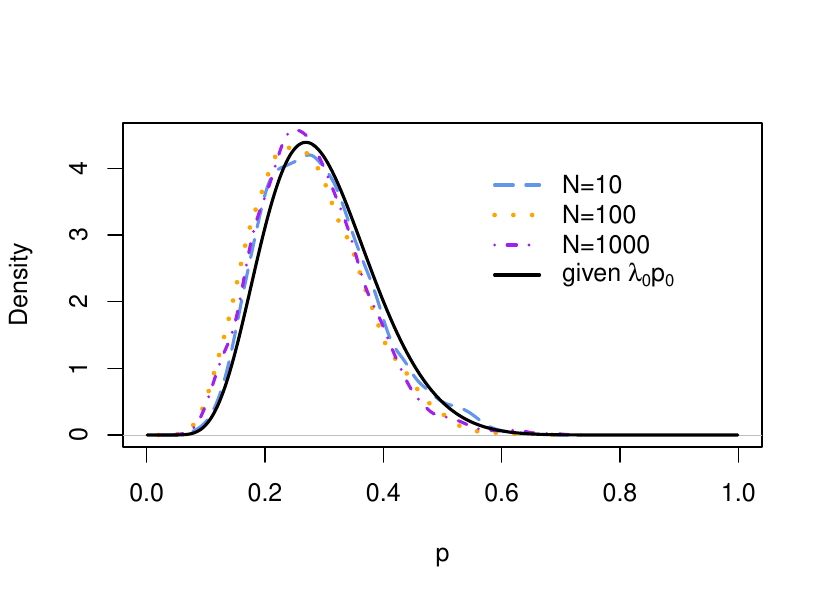} & \includegraphics[width=6.5cm]{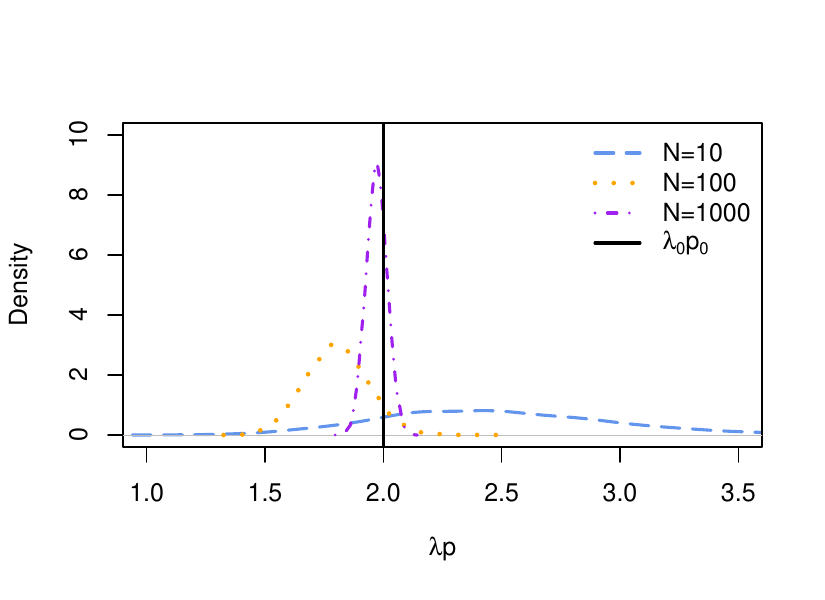} \\
    (a) & (b)
\end{tabular}
    \caption{The posterior distribution of $\lambda_{0}p_{0}$ concentrates as the amount of observational data increases, but the posterior distribution of $p_{0}$ has some amount of uncertainty that cannot be eliminated by observational data; (a) Kernel density estimates of posterior samples of $p_{0}$, conditioned on observed datasets of size $N$, with a solid line representing $p(p_{0} | \lambda_{0}p_{0})$,  (b) Kernel density estimates of posterior samples of $\lambda_{0}p_{0}$, conditioned on observed datasets of size $N$}
\label{fig:posterior_evolution_toy}
\end{center}
\end{figure}

Let us pause to examine what this estimation method can and cannot do in terms of resolving the central identifiability issue introduced in Sections \ref{sec:intro} and \ref{sec:related_work}. First, consider a toy example in which $z_{1}, \dots, z_{N} \overset{iid}{\sim} \textrm{Poisson}(\lambda_{0})$ are latent and $x_{i}\sim \textrm{Binom}(z_{i}, p_{0}), \, i=1, \dots, N$ are observed. The observed dataset identifies the product $\lambda_{0}p_{0}$, while only the prior distributions provide information about $\lambda_{0}$ and $p_{0}$ individually. The posterior distribution of the product $\lambda_{0}p_{0}$ concentrates as the number of observations increases. The posterior distributions of $\lambda_{0}$ and $p_{0}$, however, do not become arbitrarily concentrated in the infinite data limit, but rather approach $p(\lambda_{0} | \lambda_{0}p_{0})$ and $p(p_{0} | \lambda_{0}p_{0})$ respectively, as demonstrated in Figure \ref{fig:posterior_evolution_toy}.

Furthermore, the posterior distribution $p(\lambda_{0}|x_{1}, \dots, x_{N})$ will not necessarily be centered at the true value $\lambda_{0}$ (and likewise for $p_{0}$); the data provide information about the product $\lambda_{0}p_{0}$, but conditional on $\lambda_{0}p_{0}$, posterior estimates of $\lambda_{0}$ and $p_{0}$ are determined by their respective prior distributions.

\begin{figure}[H]
\begin{center}
\begin{tabular}{c c}
    \includegraphics[width=6.5cm]{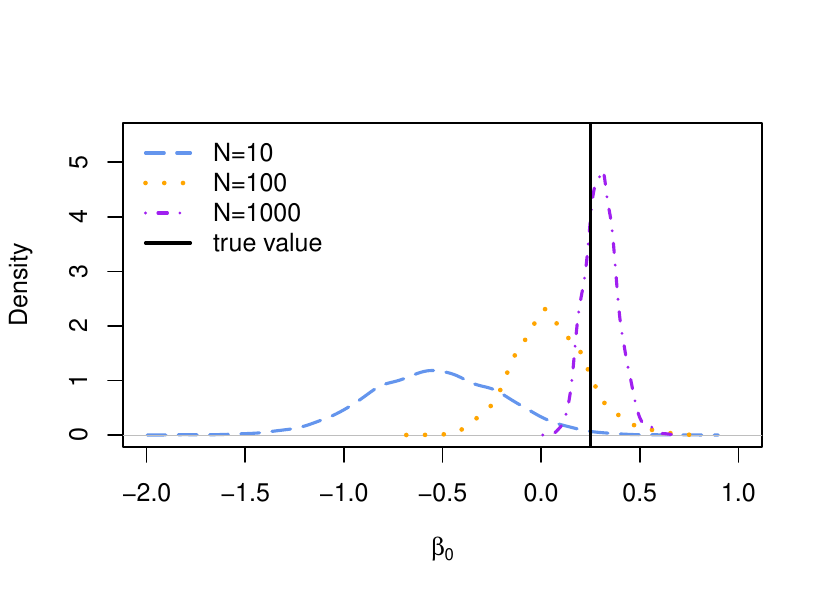} & \includegraphics[width=6.5cm]{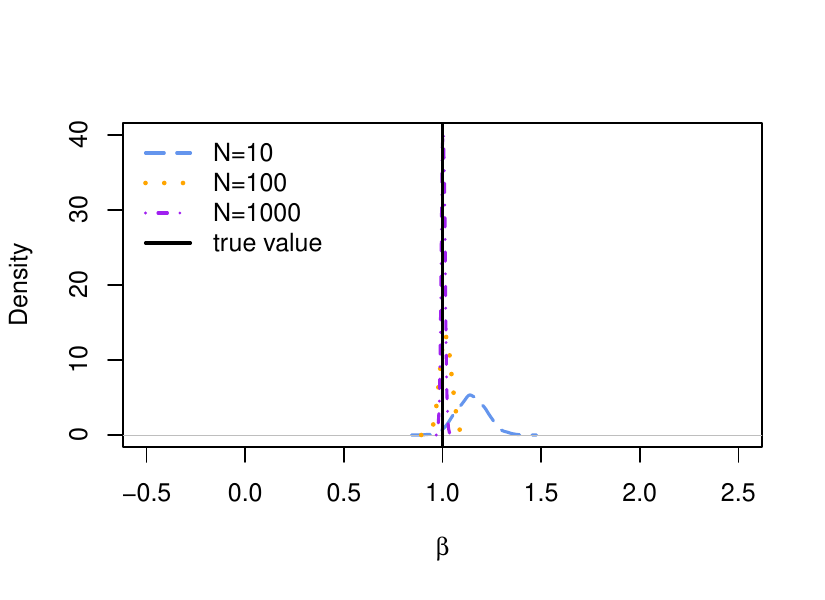} \\
    (a) & (b)
\end{tabular}
    \caption{Additional observational data is more effective at reducing posterior uncertainty about the slope coefficient $\beta$ than about the intercept ; (a) Kernel density estimates of posterior samples of $\beta_{0}$, conditioned on observed datasets of size $N$ (b) Kernel density estimates of posterior samples of $\beta_{0}$, conditioned on observed datasets of size $N$, displayed on x-axis interval of equal width as left-hand plot}
\label{fig:posterior_evolution_covars}
\end{center}
\end{figure}

A similar relationship emerges when the toy model is extended to include covariates for $\lambda$ and $p$. Let $\log(\lambda_{i}) = \beta_{0} + \beta \textrm{v}_{i}$ and $\log(\frac{p_{i}}{1-p_{i}})= \alpha_{0} + \alpha \textrm{w}_{i}$ for $\textrm{v}_{i}, \textrm{w}_{i} \in \mathbb{R}$. For $i=1, \dots, N$ latent counts are generated as $z_{i} \sim \textrm{Poisson}(\lambda_{i})$, and we observe the reported counts $x_{i} \sim \textrm{Binom}(z_{i}, p_{i})$. Performing posterior inference on data simulated from this model reveals that increasing the amount of observational data is more effective at reducing posterior uncertainty about the slope parameters $\beta$ and $\alpha$ than about the intercepts $\beta_{0}$ and $\alpha_{0}$. Figure \ref{fig:posterior_evolution_covars} illustrates the relative difficulty of learning one of the intercept parameters from data.

In light of these dynamics, Appendix \ref{app:sensitivity} explores sensitivity to the prior information elicited from NCVS estimates. While some reasons for not reporting a sexual assault are less relevant in a survey setting (such as fear of retaliation or lack of faith that the perpetrator will be held accountable), it is nevertheless plausible that some NCVS respondents choose not to disclose their sexual assault victimization. In line with the discussion in this section, inferences about the role of covariates are relatively more stable if the extent of underreporting is larger than estimated by NCVS, while inferences about the true overall incidence and reporting rate exhibit more sensitivity.

%%%%%%%%%%%%%%%%%%%%%%%%%%%%%%%%%%%%%%%%%%%%%%
%%%%%%%%%%%%%%%%%%%%%%%%%%%%%%%%%%%%%%%%%%%%%%

\section{Results}
\label{sec:results}

\subsection{Role of Covariates}
\label{sec:model_estimates}

Sampling from the model in Section \ref{sec:model}  yields the posterior means and quartiles given in Table \ref{table:posterior_summary} below.
%\begin{center}
\begin{table}[H]
\caption{Posterior means and quartiles of latent variables}
\begin{tabular}{ c | c | c | c | c | c }  
\hline
Variable & Mean & 25\% & Median & 75\% & R hat\dag  \\
\hline
$\beta_{1}$ & 0.82 & 0.81 & 0.82 & 0.84 & 1.00 \\
$\beta_{2}$ & -4.05 & -4.53 & -4.05 & -3.56 & 1.00 \\
$\beta_{0,1}$ & -4.54 & -4.67 & -4.54 & -4.42 & 1.00 \\
$\beta_{0,2}$ & -4.40 & -4.53 & -4.40 & -4.28 & 1.00 \\
$\beta_{0,3}$ & -4.32 & -4.45 & -4.32 & -4.21 & 1.00 \\
$\alpha_0$ & -1.43 & -1.50 & -1.43 & -1.36 & 1.00 \\
$\alpha_1$ & -1.97 & -2.06 & -1.97 & -1.88 & 1.00 \\
$\alpha_2$ & -2.48 & -2.73 & -2.48 & -2.23 & 1.00 \\
$\alpha_3$ & 0.49 & 0.26 & 0.49 & 0.73 & 1.00 \\
$\alpha_4$ & -3.00 & -3.18 & -3.00 & -2.83 &  1.00\\
\hline
\end{tabular}
\begin{tabular}{c}
\dag R hat, also known as the potential scale reduction factor, is a convergence diagnostic\\ proposed in \citet*{gelman1992inference}
\end{tabular}
\label{table:posterior_summary}
\end{table}
%\end{center}
\par  The posterior distribution for $\beta_1$ is concentrated around 0.82, which corresponds to sub-linear growth in the expected number of assaults as a function of number of students. The proposed the power law relationship between student population and expected number of assaults \eqref{eq:powerlaw} implies a similar relationship between student population and the expected per-capita rate of assaults, $\frac{z_{ij}}{\textrm{student population}_{ij}}$:

\begin{eqnarray}
\mathbb{E} \Bigg( \frac {z_{ij}}{\textrm{student population}_{ij}} | \boldsymbol\beta, \mathbf{v} \Bigg) &\propto& (\textrm{student population}_{ij})^{\beta_{1}-1} \nonumber .
\end{eqnarray}

Consequently, the posterior distribution suggests an inverse relationship between the size of the student population and the expected per-capita number of assaults in a given year.  With all other covariates held equal, we expect a school to have 13\% more assaults per capita than an equivalent school twice its size, and 28\% more than a school 4 times its size. Similarly, because the posterior distribution suggests $\beta_1<1$, the variance of the per capita number of assaults, $\textrm{var} \big ( \frac{z_{ij}} {\textrm{student population}_{ij}}|\boldsymbol{\beta}, \mathbf{v} \big)$, scales inversely with the number of students. With other covariates held constant, the expected number of assaults per capita is more variable at small schools than at large ones.

\par Posterior distributions for the intercepts $\beta_{0,1}$, $\beta_{0,2}$, and $\beta_{0,3}$ are similar to each other. Results are mildly suggestive of higher sexual assault incidence in rural areas than in cities, but do not reveal a substantial difference in expected number of assaults based on degree of campus urbanization. The negative values for $\beta_2$ suggest that a gender imbalance in a school's student body is associated with lower expected incidence of sexual assault, while $\alpha_{3}>0$ suggests that a higher proportion of female students corresponds to a higher reporting probability.
 Conversely, a higher proportion of students receiving Pell grants (a proxy for lower socioeconomic status) appears associated with a lower probability of reporting. Negative estimates for $\alpha_1$ and $\alpha_2$ signal lower expected reporting probabilities for junior colleges and institutions of religious instruction.
%

%%%%%%%%%%%%%%%%%%%%%%%%%%%%%%%%%%%%%%%%%%%%%%

\subsection{Systemwide Results}
\label{sec:systemwide}
Drawing posterior samples of the true number of assaults at each school induces a distribution over the total number of assaults across schools each year. The posterior median for total assaults ranges from 2.6 to 2.8 per 1000 persons over the 2014-2019 time period. Although the median estimated incidence is higher in later years compared to 2014, posterior uncertainty is large enough that the true trend in incidence could conceivably be flat or even increasing, as can be seen in Figure \ref{fig:posterior_incidence_yearly} below.

\begin{figure}[H]
\includegraphics[width=8cm]{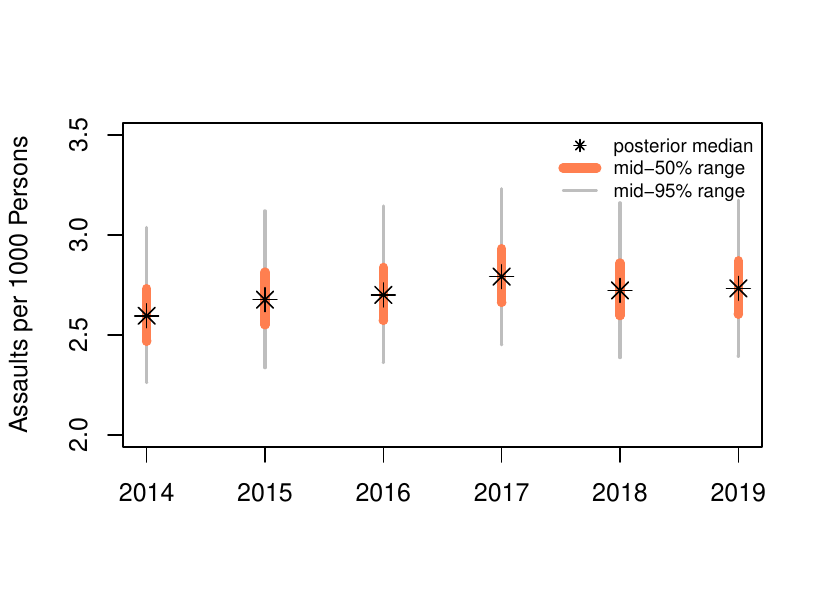}
\caption{Posterior distribution of total incidence per 1000 persons in the college population does not exhibit a clear pattern of year-over-year change.}
\label{fig:posterior_incidence_yearly}
\end{figure}

\begin{figure}[H]
\includegraphics[width=8cm]{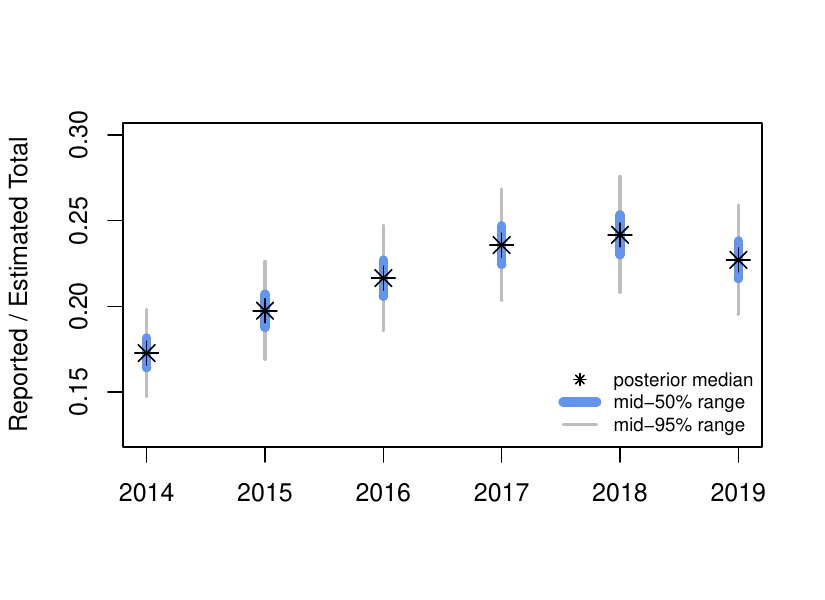}
\caption{Posterior estimates of the overall reporting rate in the college population exhibit an increasing trend from 2014 to 2018.}
\label{fig:posterior_reporting_yearly}
\end{figure}

Posterior estimates of the true reporting rate, however, show more movement over time. Figure \ref{fig:posterior_reporting_yearly} depicts posterior summaries for the actual number of reported assaults across schools as a fraction of the estimated true number of assaults. The posterior median reporting rate was lowest in 2014, at 17.3\%, and highest in 2018 at 24.2\%, with reporting rates trending upward over those years.

Taken together, these two results suggest that the increase in total sexual assaults reported nationwide over the 2014-2019 period is more likely attributable to an increase in reporting rates than an increase in the true number of assaults that occurred.

\subsection{School-Level Results}
\begin{figure}[H]
\begin{center}
\begin{tabular}{c c}
    \includegraphics[width=6.5cm]{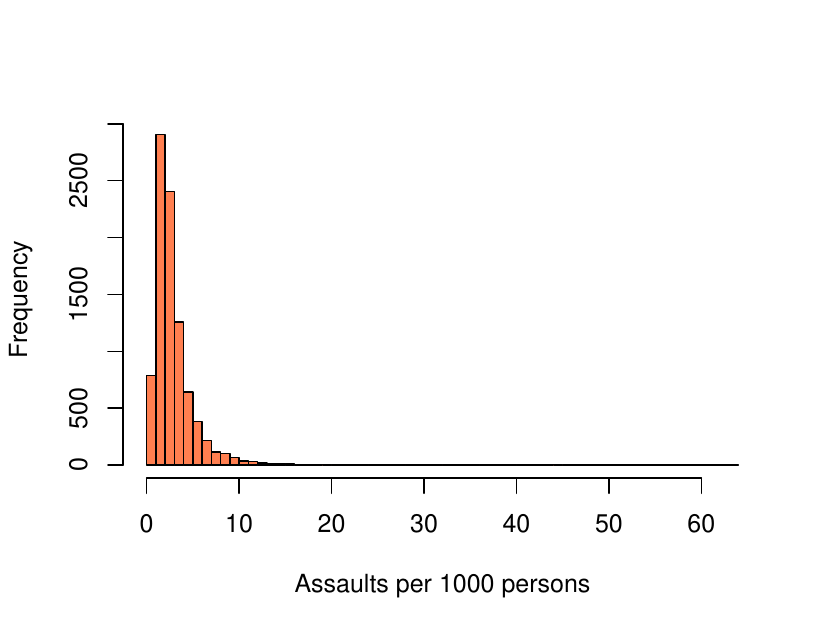} & \includegraphics[width=6.5cm]{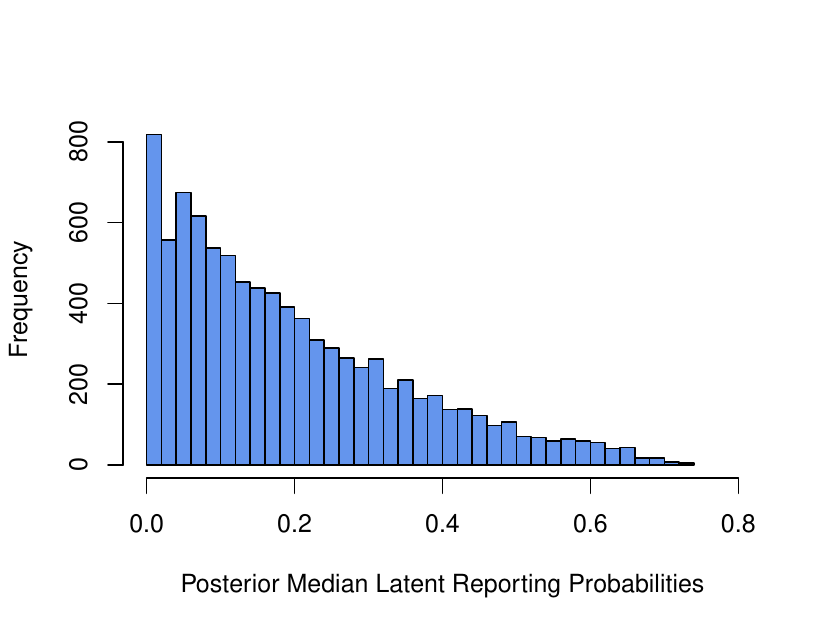} \\
    (a) & (b)
\end{tabular}
    \caption{Per-record posterior medians for incidence and reporting rate exhibit heterogeneity; (a) Posterior medians for incidence are centered slightly below the overall population value at 2.3 per 1000, but 25\% of records have posterior median incidence greater than 3.4, and another 25\% below 1.6. (b) Posterior medians for reporting probabilities are predominantly below 30\%, but for some records are as high as 74\%.}
\label{fig:posterior_medians_perrecord}
\end{center}
\end{figure}

While the systemwide increase in reported assaults may be more easily explained by an increase in reporting rates, aggregate trends are not necessarily representative of the dynamics at an individual school. Examining the posterior medians of the reporting rate $p_{ij}$ and true incidence per 1000 persons across all records in the dataset reveals considerable heterogeneity, as displayed in Figure \ref{fig:posterior_medians_perrecord}.

\par While sexual assault is susceptible to underreporting at all schools, the estimated degree of underreporting varies widely. Figure \ref{fig:posterior_medians_perrecord}(b) illustrates that some schools have extremely low reporting probabilities, while others are considerably higher than the population-level reporting rates published in the NCVS. This variation across schools has implications for how individual schools should interpret their reported crime statistics.  With only the prior belief that sexual assault is significantly underreported, one might be equally inclined to attribute an increase in reported assaults at any school to an increase in the reporting probabilities. However, this may lead to flawed conclusions.

\par Suppose that for a particular school, the true number of assaults in a future year, included in the dataset, remains the same as the true number of assaults in 2019. With an unchanged number of true assaults, how large of a change in the reported number of assaults is plausible due to variation in the reporting probability alone? For school $i$, the predictive distribution for a future year's number of reported assaults can be sampled as follows.

\begin{enumerate}[\indent(1)]
    \item Draw $(z_{i,2019}^{(s)}, \alpha_{0}^{(s)}, \boldsymbol{\alpha}^{(s)}, \gamma_{i}^{(s)})$ from model's posterior distribution
    \item Draw $\delta^{new} \sim N(0,0.5)$
    \item Set $p^{new} = \textrm{invlogit}(\alpha_{0}^{(s)} +  \langle \boldsymbol{\alpha}^{(s)}, \mathbf{w}_{i,2019} \rangle + \gamma_{i}^{(s)} + \delta^{new}) $
    \item Draw $x^{new} \sim \textrm{Binom}(z_{i,2019}, p^{new})$
\end{enumerate}
\vspace{12pt}

\begin{figure}[H]
\begin{center}
\begin{tabular}{c c}
    \includegraphics[width=6.5cm]{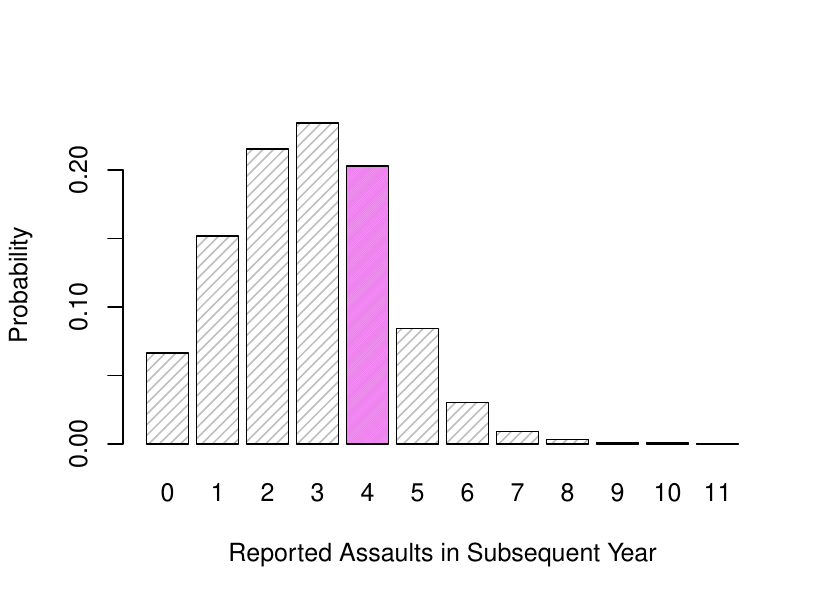} & \includegraphics[width=6.5cm]{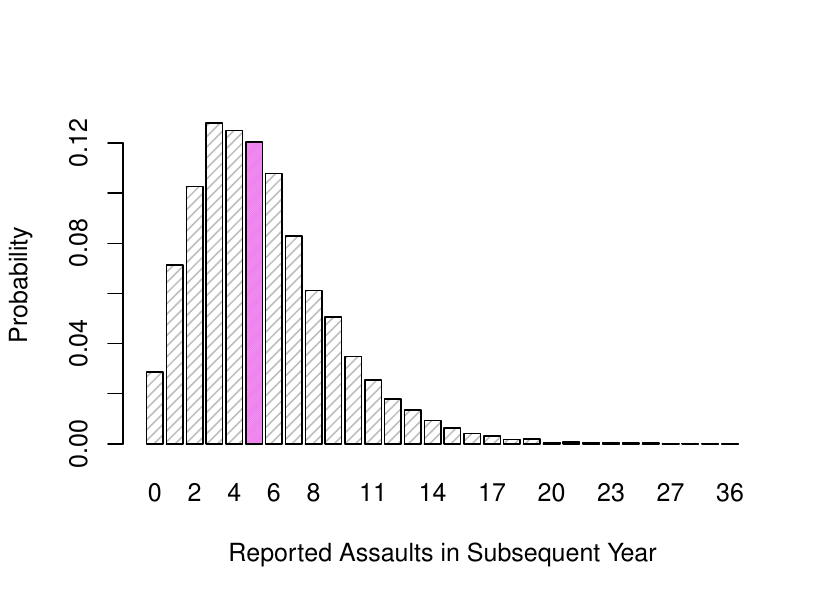} \\
    (a) & (b)
\end{tabular}
    \caption{Variation in reporting probability alone is less likely to produce an increased number of reported assaults in subsequent year at (a) Randolph College than at (b) University of South Carolina- Columbia. Solid-colored bar corresponds to number of reported assaults in 2019.}
\label{fig:constant_z1}
\end{center}
\end{figure}

Randolph College and the University of South Carolina-Columbia (USCC) had a similar number of reported assaults in 2019 (4 and 5 respectively). However, Randolph College's estimated reporting rate was considerably higher, with a posterior median of 53\% compared to only 14\% at USCC. Assuming that the true number of assaults is held constant, the predictive distributions for the number of reported assaults are shown in Figure \ref{fig:constant_z1}.

Under this scenario, there is a roughly 13\% chance of an increased number of reported assaults at Randolph College in a subsequent year due to variation in the reporting rate, and less than a 1\% chance of the reported number to double to 8 or more. At USCC, variation in reporting rate has a comparatively larger chance of producing an increased number of reported assaults (42\%), and a 12\% chance of doubling the number of reported assaults to 10 or more.

\par
\begin{figure}[H]
\begin{center}
\begin{tabular}{c c}
    \includegraphics[width=6.5cm]{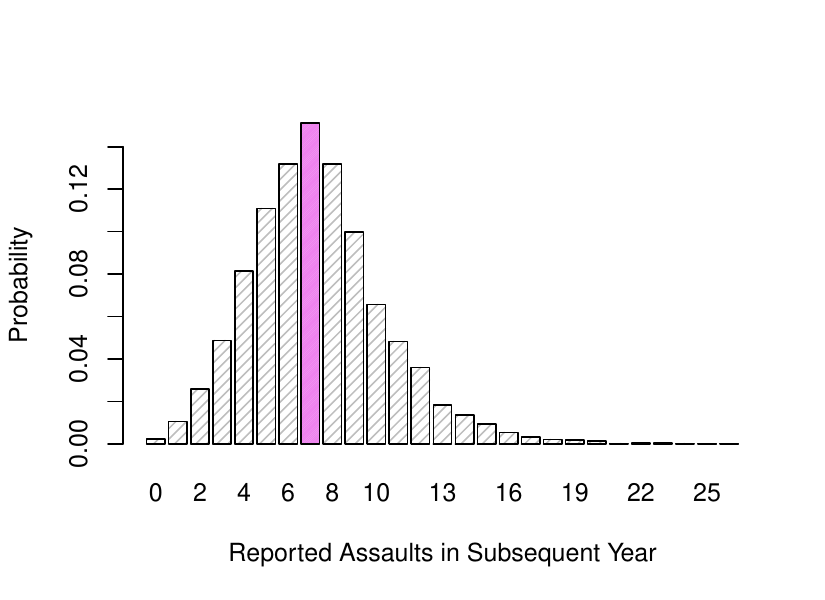} & \includegraphics[width=6.5cm]{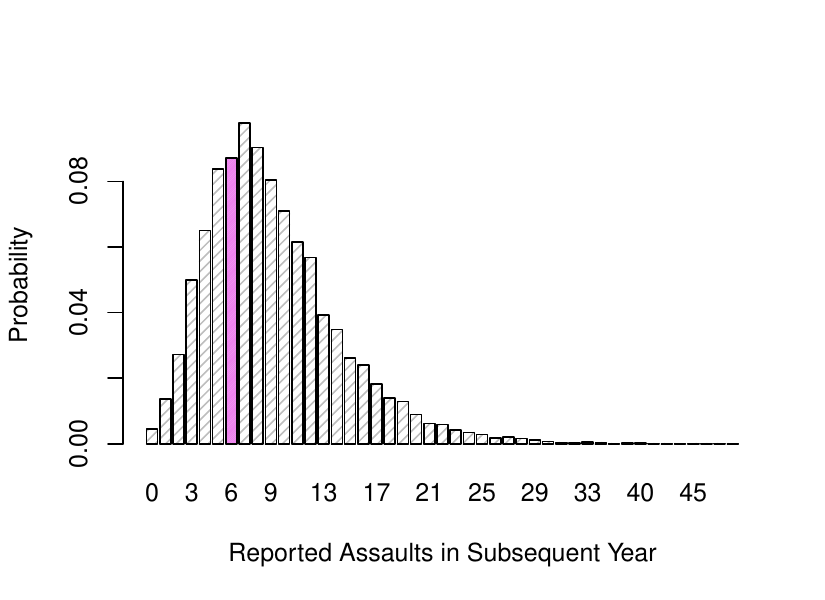} \\
    (a) & (b)
\end{tabular}
    \caption{Without a change in the true number of assaults, the number of reported assaults has a 44\% chance of increasing in a subsequent year at (a) Ursinus College, and a 67\% chance at (b) Temple University.}
\label{fig:constant_z2}
\end{center}
\end{figure}

 Ursinus College and Temple University provide another illustration of such school-level differences. Of the two, Ursinus had a much higher estimated reporting probability in 2019, and changes in reporting probability are comparatively less likely to drive an increase in reported assaults at Ursinus compared to Temple, as can be seen from their respective predictive distributions in Figure \ref{fig:constant_z2}.

These disparities underscore the opaque nature of reported campus sexual assault data and the potential pitfalls of a one-size-fits-all approach to interpreting school-level reports over time.

%%%%%%%%%%%%%%%%%%%%%%%%%%%%%%%%%%%%%%%%%%%%%%
%%%%%%%%%%%%%%%%%%%%%%%%%%%%%%%%%%%%%%%%%%%%%%

\section{Discussion}
\label{sec:conclusion}
\par Colleges and universities prioritize reducing the incidence of sexual assault on their campuses, but underreporting diminishes the interpretability of their published sexual assault incidence data. This paper constructs a generative model of underreported campus sexual assault data which allows for estimation of true incidence and reporting rates. Fitting this model reveals that lower socioeconomic status, as measured by the percentage of a school's undergraduates receiving Pell grants, appears associated with a lower probability of reporting. Status as a junior college or institution of religious instruction appear similarly associated with lower reporting probability, while gender imbalance in a school's student body is associated with lower true incidence of assault.

\par For the overall college population, estimated incidence of sexual assault remained fairly stable over 2014-2019, while reporting probabilities increased. Results at individual schools, however, varied widely. Estimates of school-level incidence and reporting probabilities may help university officials, for instance Title IX coordinators, assess the effectiveness of initiatives to reduce incidence of sexual assault or to familiarize students with reporting resources, and decide where to make changes and improvements moving forward.
\par One avenue for future work concerns repeat victimizations of the same individual. High-frequency series victimizations, for instance in the case of patterns of intimate partner violence, can produce extreme values in the reported number of assaults. As an example, in 2017 the University of Nebraska-Lincoln received 104 reports of sexual assault corresponding to the same victim and perpetrator. Such extreme outliers are not easily accommodated, and in particular may violate the modeling assumption that the reporting decisions for each assault are independent (conditional on reporting probability). Violations of the independence assumption can degrade the quality of posterior inferences- see Appendix \ref{app:independence_violation} for further details. More broadly, the Bureau of Justice Statistics has acknowledged the difficulty of accurately incorporating series victimization in its national crime estimates \citep{repeatvictimizations}.

%%%%%%%%%%%%%%%%%%%%%%%%%%%%%%%%%%%%%%%%%%%%%%
%% Appendix---Please move all appendices to %%
%% a Supplementary file.                    %%
%%%%%%%%%%%%%%%%%%%%%%%%%%%%%%%%%%%%%%%%%%%%%%
%% Support information, if any,             %%
%% should be provided in the                %%
%% Acknowledgements section.                %%
%%%%%%%%%%%%%%%%%%%%%%%%%%%%%%%%%%%%%%%%%%%%%%
\begin{acks}[Acknowledgments]
 The authors would like to thank the referees, Associate Editor and the Editor for their constructive comments and suggestions that improved the quality of this paper.\\
 David M. Blei is also affiliated with the Department of Computer Science at Columbia University.

\end{acks}
%%%%%%%%%%%%%%%%%%%%%%%%%%%%%%%%%%%%%%%%%%%%%%
%% Funding information, if any,             %%
%% should be provided in the                %%
%% funding section.                         %%
%%%%%%%%%%%%%%%%%%%%%%%%%%%%%%%%%%%%%%%%%%%%%%
\begin{funding}
% The first author was supported by ...
%
The second author is supported by NSF IIS-2127869, NSF DMS-2311108, ONR N000142412243, and the Simons Foundation.
\end{funding}

%%%%%%%%%%%%%%%%%%%%%%%%%%%%%%%%%%%%%%%%%%%%%%
%% Supplementary Material, including data   %%
%% sets and code, should be provided in     %%
%% {supplement} environment with title      %%
%% and short description. It cannot be      %%
%% available exclusively as external link.  %%
%% All Supplementary Material must be       %%
%% available to the reader on Project       %%
%% Euclid with the published article.       %%
%%%%%%%%%%%%%%%%%%%%%%%%%%%%%%%%%%%%%%%%%%%%%%
\begin{supplement}
\stitle{Appendices}
\sdescription{This file contains Appendices A-E to the main text.}
\end{supplement}
\begin{supplement}
\stitle{Code and Data}
\sdescription{This file contains R and Stan implementations of methods from the main text, along with preprocessed data.}
\end{supplement}

%%  imsart-nameyear.bst  will be used to                   %%
%%  MR numbers will be added by VTeX.                      %%
%%                                                         %%
%%  Use \cite{...} to cite references in text.             %%
%%                                                         %%
%%%%%%%%%%%%%%%%%%%%%%%%%%%%%%%%%%%%%%%%%%%%%%%%%%%%%%%%%%%%%

%% if your bibliography is in bibtex format, uncomment commands:
\bibliographystyle{imsart-nameyear} % Style BST file
\bibliography{references}       % Bibliography file (usually '*.bib')

%% or include bibliography directly:
% \begin{thebibliography}{}
% \bibitem[\protect\citeauthoryear{???}{???}]{b1}
% \end{thebibliography}

%%%%%%%%%%%%%%%%%%%%%%%%%%%%%%%%%%%%%%%%%%%%%%
%% Appendix---Please move all appendices to %%
%% a Supplementary file.                    %%
%%%%%%%%%%%%%%%%%%%%%%%%%%%%%%%%%%%%%%%%%%%%%%
%% Support information, if any,             %%
%% should be provided in the                %%
%% Acknowledgements section.                %%
%%%%%%%%%%%%%%%%%%%%%%%%%%%%%%%%%%%%%%%%%%%%%%

\begin{appendix}

\newpage
\section{Dataset Details}
\label{app:data_description}

The foregoing analysis pertains to postsecondary schools in the United States which grant academic degrees, reported a positive number of in-person students, and which submitted campus security reports under the Clery Act during the 2014-2019 period. A limited number of US colleges and universities are not subject to these reporting requirements.
\par Data for number of reported assaults was furnished by the US Department of Education Office of Postsecondary Education, Campus Safety and Security database. All other data is sourced from the Integrated Postsecondary Education Data System (IPEDS), maintained by the National Center for Education Statistics (NCES), a division of the US Department of Education.

\textbf{Number of reported assaults}- For each school in each year, this is the number of sexual assaults disclosed in the campus security report corresponding to that calendar year.

\textbf{Student population}- For each school in each year, this consists of the total number of students enrolled in the Fall, minus the number of students enrolled in 100\% distance learning programs.

\textbf{Percent female}- For each school in each year, this is the percent of the student body enrolled in the fall who were classified as female (`male' and `female' are currently the only gender identity options in the reporting system).

\textbf{Degree of urbanization}- Degree of urbanization is categorized on a 12-point scale defined by the NCES ( \url{https://nces.ed.gov/programs/edge/docs/LOCALE_CLASSIFICATIONS.pdf}) For the purpose of the foregoing analysis, this scale was collapsed into 3 categories corresponding to `urban', `suburban', and `rural'.

\textbf{Pell grant recipients}- For each school in each year, this is the percent of undergraduate students who received Pell grants.

\textbf{Associate degree only}- For each school, this is an indicator for whether the school offers only associate degrees. This equals zero for schools offering bachelor, master, or doctoral degrees.

\textbf{Religious instruction}- For each school, this is an indicator for whether the school is substantially engaged in religious instruction. Theological seminaries are one such example. This is a manually curated subset of schools designated as ``Private not-for-profit (religious affiliation)" in IPEDS. This variable is included in the Supplementary Material.

\vspace{24pt}

\textbf{Modifications and exclusions}
\begin{itemize}
    \item Michigan State University is excluded from the data set due to notable unreliability of its reporting systems, for which the school was fined over \$4 million by the US Department of Education in 2019.
    \item Northern Oklahoma College (NOC) is a community college. One of its campuses is co-located with the main Stillwater, OK campus of Oklahoma State University (OSU), a large flagship public university. Beginning in 2019, campus crime statistics for NOC include crimes on the shared OSU Stillwater campus. Consequently, these assaults are double counted in the dataset. For 2019, we attribute these assaults to OSU only and do not duplicate them for NOC.
    \item For Ohio State University (main campus), the reported number of assaults in 2018 and 2019 included assaults perpetrated by former Ohio State physician Richard Strauss. These assaults occurred during his 20-year employment at Ohio State, from 1978 to 1998. We exclude these assaults from our analysis (30 in 2018 and 97 in 2019), as they occurred long before the relevant reporting period. (Note: Strauss died in 2005.)
    \item In 2017 University of Nebraska-Lincoln had a total of 119 reported assaults. Of those 119 reported assaults, 104 corresponded to the same victim-perpetrator pair. This type of high-frequency serial victimization is beyond the scope of what the current model can handle; we defer this consideration to future work. Instead, we record one assault for the repeat victim/perpetrator pair.
    \item In 2017 Wells College had 488 students and 29 reported assaults. Of those 29 reported assaults, 28 corresponded to the same victim and perpetrator \citep{axelson2019}. This type of high-frequency serial victimization is beyond the scope of what the current model can handle; we defer this consideration to future work. Instead, we record one assault for the repeat victim/perpetrator pair.
    \item In 2015 Genesee Community College had 20 reported assaults, all 20 of which corresponded to the same victim and perpetrator \citep{goodman2017}. This type of high-frequency serial victimization is beyond the scope of what the current model can handle; we defer this consideration to future work. Instead, we record one assault for the repeat victim/perpetrator pair.
\end{itemize}

%%%%%%%%%%%%%%%%%%%%%%%%%%%%%%%%%%%%%%%%
%%%%%%%%%%%%%%%%%%%%%%%%%%%%%%%%%%%%%%%%

\newpage

\section{Derivations}
\label{app:derivations}

\begin{claim}
$z| \lambda, p \sim \textrm{Poisson}(\lambda p)$
\end{claim}
Given parameters $\lambda$ and $p$, the true number of events $z$ is independent of covariates $\mathbf{w}$ and $\mathbf{v}$, rendering the distribution of $z|\lambda, p$ equal to the distribution of $z|\lambda, p, \mathbf{w}, \mathbf{v}$. For simplicity of notation, the covariates are suppressed in the following.\newline
\vspace{12pt}
  
For an arbitrary nonnegative integer $k$, we have:
\begin{eqnarray}
\mathbb{P}(x=k|\lambda, p) &=& \sum_{j=0}^{\infty} \mathbb{P}(x=k |\lambda, p, z=j) \mathbb{P}(z=j| \lambda, p)\nonumber\\
&=& \sum_{j \geq k} \mathbb{P}(x=k| \lambda, p, z=j) \mathbb{P}(z=j| \lambda, p)\label{eq:x_support}\\
&=& \sum_{j \geq k} \binom{j}{k} p^{k} (1-p)^{j-k} \frac{e^{-\lambda} \lambda^{j}}{j!}\nonumber\\
&=& \sum_{j \geq k} \frac{j!}{k! (j-k)!} p^{k}(1-p)^{j-k} e^{-\lambda} \frac{\lambda^{j}}{j!}\nonumber\\
&=& \sum_{j \geq k} \frac{1}{k! (j-k)!} (p \lambda)^{k} ((1-p) \lambda)^{j-k} e^{- \lambda}\nonumber\\
&=& \frac{(p \lambda)^{k} e^{- \lambda}}{k!}\sum_{j \geq k} \frac{((1-p) \lambda)^{j-k}}{(j-k)!}\nonumber\\
&=& \frac{(p \lambda)^{k} e^{- \lambda}}{k!}\sum_{m=0}^{\infty} \frac{((1-p) \lambda)^{m}}{(m)!}\label{eq:change_of_var}\\
&=& \frac{(p \lambda)^{k} e^{- \lambda}}{k!} e^{(1-p) \lambda}\nonumber\\
&=& \frac{(p \lambda)^{k}}{k!} e^{-\lambda p}\nonumber.
\end{eqnarray}

In the above, \eqref{eq:x_support} holds because given $z=j$, $x$ is supported on $ \{0, 1, \dots, j \}$. Step \eqref{eq:change_of_var} utilizes a change of variables to $m \overset{\underset{\mathrm{def}}{}}{=} j-k$.

\vspace{24pt}

\begin{claim}
$u|x, \lambda, p \sim \textrm{Poisson}((1-p) \lambda)$
\end{claim}

As in Section \ref{sec:model}, consider decomposing the total number of assaults $z$ into the number of reported assaults $x$ and the number of unreported assaults $u$. When $x$ is known, the randomness in $z$ comes only from $u$, and the distribution of $z|x, \lambda, p$ is the distribution of $u|x, \lambda, p$ shifted to the right by the observed quantity $x$.

\vspace{12pt}
  
First consider the joint distribution of $u$ and $x$.
\begin{eqnarray}
    \mathbb{P}(u=l, x=k | \lambda, p) &=& \sum_{j=0}^{\infty} \mathbb{P}(u=l, x=k | z=j, \lambda, p) \mathbb{P}(z=j| \lambda, p)\nonumber\\
    &=& \mathbb{P}(u=l, x=k | z= l+k, \lambda, p) \mathbb{P}(z= l+k | \lambda, p)\label{eq:support_ux}\\
    &=& \binom{l+k}{k} p^{k} (1-p)^{l} \frac{e^{-\lambda} \lambda^{l+k}}{(l+k)!}\nonumber\\
    &=& \frac{(p \lambda)^{k} ((1-p) \lambda)^{l}}{k! l!} e^{-\lambda}\label{eq:ux_joint}
\end{eqnarray}

Step \eqref{eq:support_ux} holds because $\mathbb{P}(u=l,x=k|z, \lambda, p)$ equals zero for any value of $z$ such that $z \neq l+k$.

Then, 
\begin{equation}
    \mathbb{P}(u=l|x=k, \lambda, p) = \frac{\mathbb{P}(u=l, x=k | \lambda, p)}{\sum_{m=0}^{\infty} \mathbb{P}(u=m, x=k | \lambda, p)}.\nonumber
\end{equation}

Substituting in the expression from step \eqref{eq:ux_joint}, we have:
\begin{eqnarray}
    \mathbb{P}(u=l|x=k, \lambda, p) &=& \frac{(p \lambda)^{k} ((1-p) \lambda)^{l}e^{-\lambda} / (k! l!) }{\sum_{m=0}^{\infty} (p \lambda)^{k} ((1-p)\lambda)^{m} e^{-\lambda} / (m! k!)}\nonumber\\
    &=& \frac{((1-p) \lambda)^{l} / l!}{ \sum_{m=0}^{\infty} ((1-p)\lambda)^{m} / m!}\nonumber\\
    &=& \frac{((1-p) \lambda)^{l} / l!}{e^{(1-p)\lambda}}\nonumber\\
    &=& \frac{((1-p) \lambda)^{l} e^{-(1-p)\lambda}}{l!}.\nonumber
\end{eqnarray}
%%%%%%%%%%%%%%%%%%%%%%%%%%%%%%%%%%%%%%%%
%%%%%%%%%%%%%%%%%%%%%%%%%%%%%%%%%%%%%%%%

\newpage
\section{Pooling in Hierarchical Models}
\label{app:pooling}

\par In the model in Section \ref{sec:model}, coefficients $\mathbf{\alpha}$ and $\mathbf{\beta}$ are global, encouraging the sharing of information across schools, while school-level terms $\mathbf{\gamma}$ and $\mathbf{\varepsilon}$ capture local properties of individual schools. This model structure is a compromise between complete pooling, in which no school-level differences are permitted ($\gamma$ and $\varepsilon$ omitted), and no pooling, in which each school is modeled separately from all others, with its own coefficient vectors $\alpha_i$ and $\beta_i$. 

\par In the case of complete pooling or no pooling, the observed counts still arise from a binomial distribution,
$$x_{ij}|z_{ij}, p_{ij} \sim \textrm{Binom}(z_{ij}, p_{ij}),$$
but the true number of assaults and probability of reporting are constructed differently.

%[Could also imagine extending the model hierarchy to treat the variance of $\mathbf{\gamma}$ and $\varepsilon$ as additional latent variables to be inferred.]

\subsection{Complete Pooling}
The complete pooling scenario assumes that sexual assault reporting patterns do not systematically differ across schools, except to the extent accounted for by their covariates.

\begin{figure}[H]
\includegraphics[width=12cm]{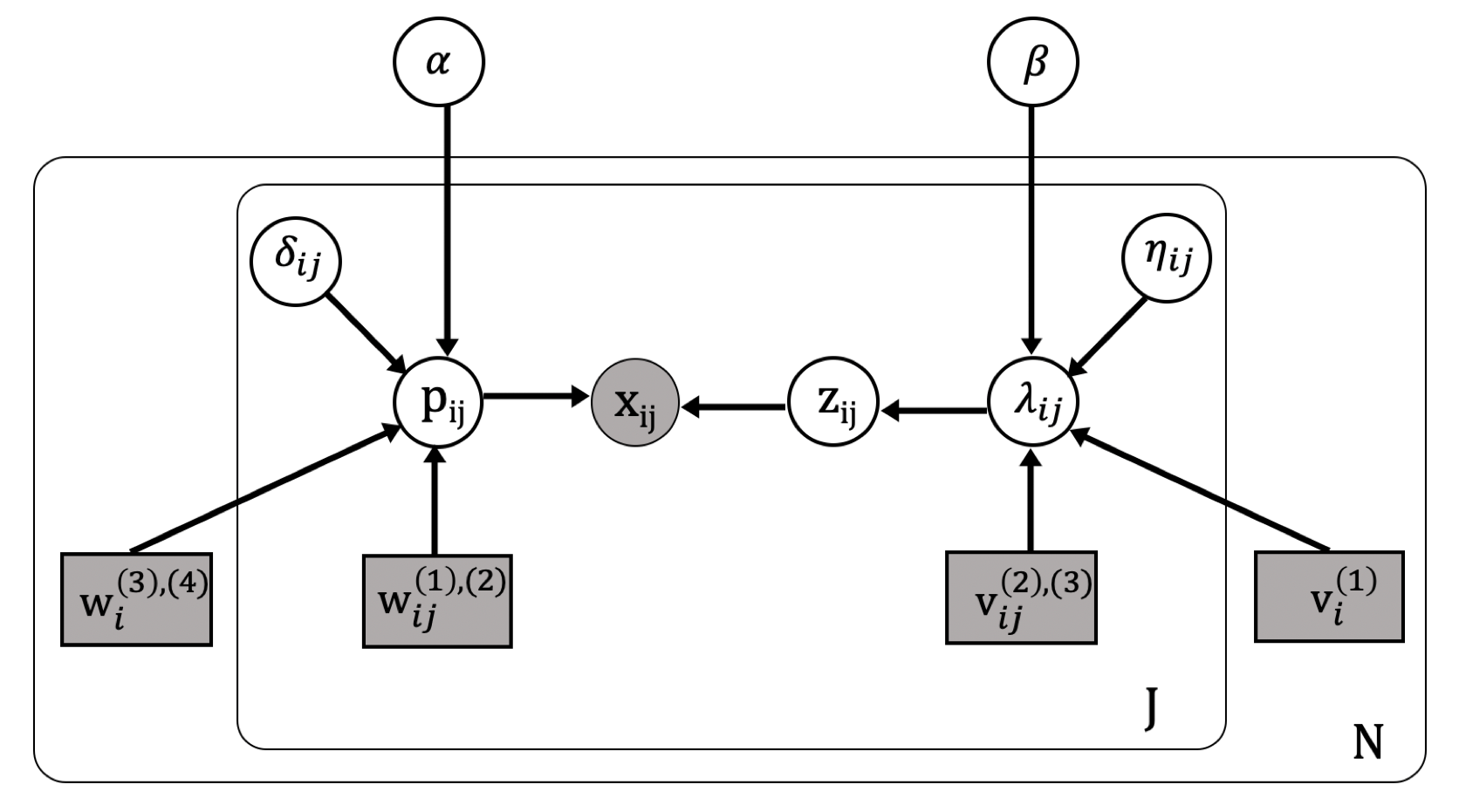}
\caption{Complete Pooling Model Diagram}
\end{figure}

The true number of assaults, $z$, follows a model similar to that of Section \ref{sec:model}, but with the school-level offset $\varepsilon_i$ omitted from Equation \ref{eq:lambda}, that is:

$$\log(\lambda_{ij}) = \beta_{0, \textrm{v}_{i}^{(1)}} + \beta_{1} \textrm{v}_{ij}^{(2)} + \beta_{2} \textrm{v}_{ij}^{(3)} + \eta_{ij}.$$

%\begin{eqnarray}
%z_{ij}|\lambda_{ij} &\sim& \textrm{Poisson}(\lambda_{ij}) \nonumber\\
%\log(\lambda_{ij}) &=& \beta_{0, \textrm{v}_{i}^{(1)}} + \beta_{1} \textrm{v}_{ij}^{(2)} + \beta_{2} \textrm{v}_{ij}^{(3)} + \eta_{ij} \nonumber\\
%\beta_{0,1}, \beta_{0,2}, \beta_{0,3}  &\sim& N(-5.5, 0.5) \nonumber\\
%\beta_{1} &\sim& N(1,0.1) \nonumber\\
%\beta_{2} &\sim& N(0,2) \nonumber\\
%\eta_{ij} &\sim& N(0, 0.2) \nonumber
%\end{eqnarray}

Likewise, the probability of reporting is modeled as in Section \ref{sec:model}, but with the school-level offset $\gamma_i$ omitted from Equation \ref{eq:p}, that is:

$$\log \Big( \frac{p_{ij}}{1-p_{ij}} \Big) = \alpha_{0}+\alpha_1 \textrm{w}_{i}^{(1)} + \alpha_{2} \textrm{w}_{i}^{(2)}+ \alpha_{3} \textrm{w}_{ij}^{(3)} + \alpha_{4} \textrm{w}_{ij}^{(4)} + \delta_{ij}.$$

%\begin{eqnarray}
%\log \Big( \frac{p_{ij}}{1-p_{ij}} \Big) &=& \alpha_{0}+\alpha_1 \textrm{w}_{i}^{(1)} + \alpha_{2} \textrm{w}_{i}^{(2)}+ \alpha_{3} \textrm{w}_{ij}^{(3)} + \alpha_{4} \textrm{w}_{ij}^{(4)} + \delta_{ij} \nonumber\\
%\alpha_{0} &\sim& N(-1.25, 0.5)\nonumber\\
%\alpha_{1}, \alpha_{2}, \alpha_{4} &\sim& N(0,1)\nonumber\\
%\alpha_{3} &\sim& N(0,2)\nonumber\\
%\delta_{ij} &\sim& N(0, 0.5)\nonumber
%\end{eqnarray}

\subsection{No Pooling}
In the no-pooling scenario, each school is modeled separately. 

\begin{figure}[H]
\includegraphics[width=9cm]{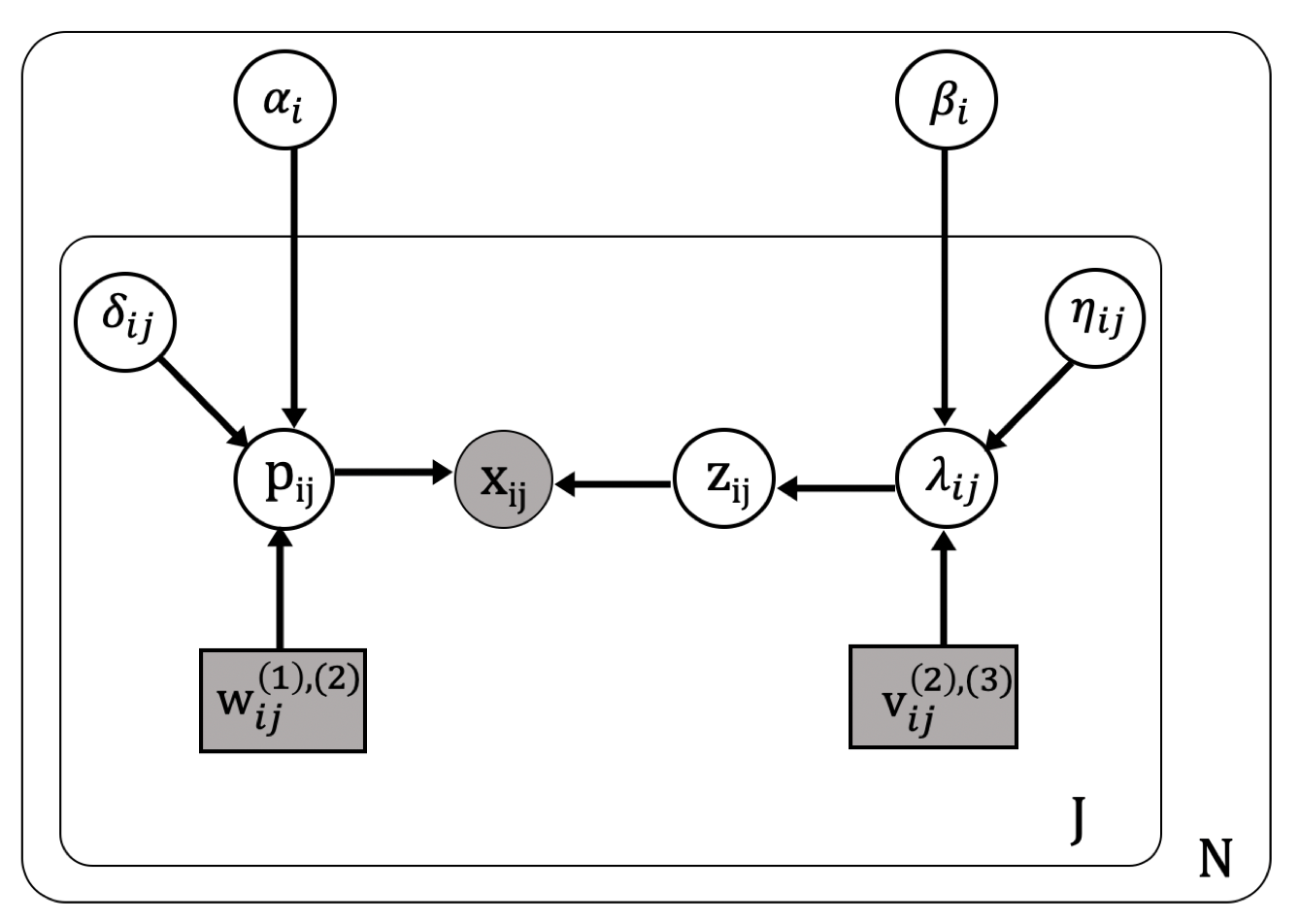}
\caption{No-Pooling Model Diagram}
\end{figure}

%The true number of assaults is modeled as follows:
%\begin{eqnarray}
%z_{ij}|\lambda_{ij} &\sim& \textrm{Poisson}(\lambda_{ij}) \nonumber\\
%\log(\lambda_{ij}) &=& \beta_{0,i} + \beta_{1,i} \textrm{v}_{ij}^{(2)} + \beta_{2,i} \textrm{v}_{ij}^{(3)} + \eta_{ij} \nonumber\\
%\beta_{0, i}  &\overset{iid}{\sim}& N(-5.5, 0.5) \nonumber\\
%\beta_{1, i} &\overset{iid}{\sim}& N(1,0.1) \nonumber\\
%\beta_{2, i} &\overset{iid}{\sim}& N(0,2) \nonumber\\
%\eta_{ij} &\overset{iid}{\sim}& N(0, 0.2). \nonumber
%\end{eqnarray}

This model has a separate set of coefficients $\beta_{0,i}, \beta_{1,i}, \beta_{2,i}$ for each school $i$. Each model has a single intercept rather than separate intercepts for urban, suburban, and rural locations, as each school has a single location and does not allow for this distinction. The school-level offset $\varepsilon_{i}$ is omitted; when modeling a single school, the dataset does not provide any information that distinguishes $\varepsilon_{i}$ from the intercept $\beta_{0,i}$. When modeling $z$, we thus replace Equation \ref{eq:lambda} from the original partial pooling model with:

$$\log(\lambda_{ij}) = \beta_{0,i} + \beta_{1,i} \textrm{v}_{ij}^{(2)} + \beta_{2,i} \textrm{v}_{ij}^{(3)} + \eta_{ij}.$$

%The reporting probabilities are modeled as follows:
%\begin{eqnarray}
%\textrm{log} \Big( \frac{p_{ij}}{1-p_{ij}} \Big) &=& \alpha_{0,i} + \alpha_{3,i} \textrm{w}_{ij}^{(3)} + \alpha_{4,i} \textrm{w}_{ij}^{(4)} + \delta_{ij} \nonumber\\
%\alpha_{0,i} &\overset{iid}{\sim}& N(-1.25, 0.5)\nonumber\\
%\alpha_{4,i} &\overset{iid}{\sim}& N(0,1)\nonumber\\
%\alpha_{3,i} &\overset{iid}{\sim}& N(0,2)\nonumber\\
%\delta_{ij} &\overset{iid}{\sim}& N(0, 0.5).\nonumber
%\end{eqnarray}

Each school $i$ has a separate set of coefficients $\alpha_{0,i}, \alpha_{3,i}, \alpha_{4,i}$. As above, the school-level offset $\gamma_{i}$ is omitted, as the dataset does not distinguish it from the intercept $\alpha_{0,i}$. Coefficients $\alpha_1$ and $\alpha_2$ are omitted. These coefficients correspond whether a school is a junior college, and whether a school is primarily engaged in religious instruction, respectively. As these are fixed characteristics of a school, the data for a single school do not provide information to help distinguish these coefficients from the intercept $\alpha_{0,i}$. When modeling $z$, we thus replace Equation \ref{eq:p} from the original partial pooling model with:

$$\log \Big( \frac{p_{ij}}{1-p_{ij}} \Big) = \alpha_{0,i} + \alpha_{3,i} \textrm{w}_{ij}^{(3)} + \alpha_{4,i} \textrm{w}_{ij}^{(4)} + \delta_{ij}.$$

\subsection{Comparison}

Roughly 20\% of records in the campus sexual assault dataset
were held out, and posterior inference for the partial pooling, complete pooling, and no pooling models was carried out on the remaining 80\% of records. To compare the three models, we approximate the predictive likelihood on the held-out data. Denoting a model's collection of latent variables as $\boldsymbol\theta$ and suppressing covariates, this quantity is 
$$p(x^{held-out}| x^{train}) = \int p(x^{held-out}|\boldsymbol\theta) p(\boldsymbol\theta | x^{train}) d\boldsymbol\theta. $$

\begin{figure}[H]
\begin{center}
\begin{tabular}{c c}
    \includegraphics[height=6cm]{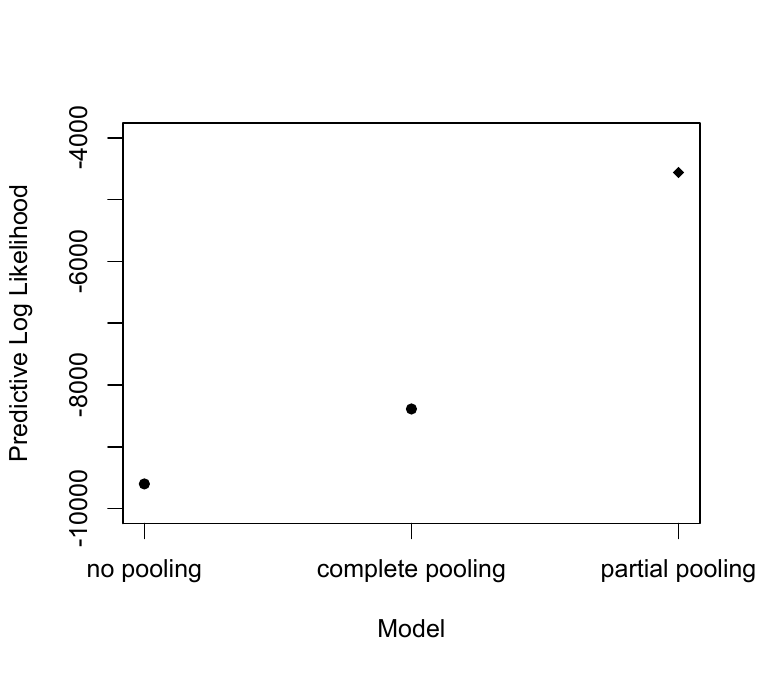} & \includegraphics[height=6cm]{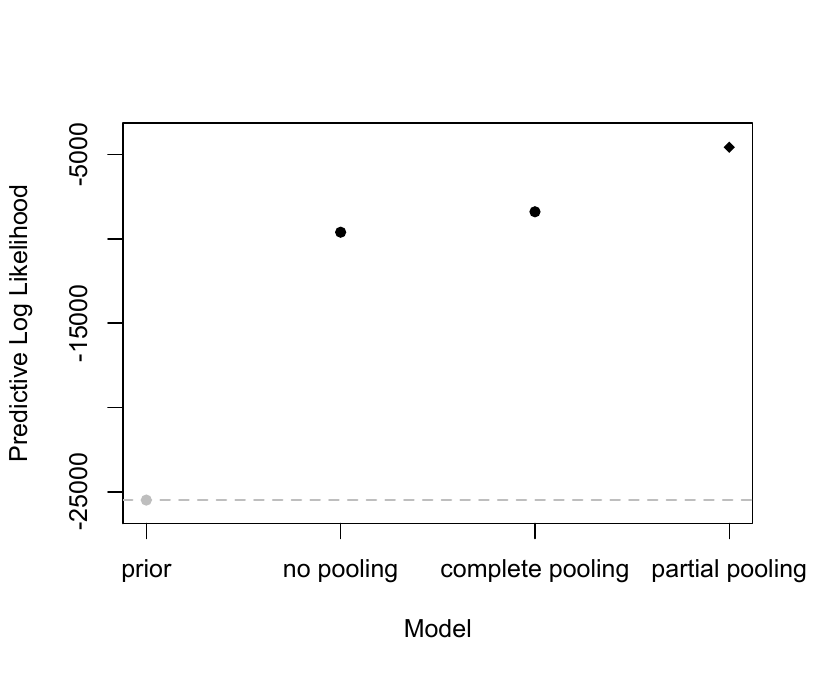} \\
    (a) & (b)
\end{tabular}
    \caption{The partial pooling model assigns higher likelihood to held-out observational data; (a) Estimated predictive log likelihood of the held-out dataset under each model (b) Left-hand side plot rescaled to include log likelihood of the held-out dataset under the prior from the partial pooling model}
\label{fig:pooling_comparison}
\end{center}
\end{figure}

Figure \ref{fig:pooling_comparison} shows that the partial pooling model assigns the highest likelihood to the held-out data. This result that favors the partial pooling model over the two alternatives, as it suggests the partial pooling model does a better job approximating the true distributioin of the data.

%%%%%%%%%%%%%%%%%%%%%%%%%%%%%%%%%%%%%%%%
%%%%%%%%%%%%%%%%%%%%%%%%%%%%%%%%%%%%%%%%

\newpage
\section{Sensitivity to Correlated Reporting}
\label{app:independence_violation}

\par The model in Section \ref{sec:model} assumes that, for a given school in a given year, each assault is reported or not reported independently of all others. In practice, this assumption could plausibly be violated. To explore sensitivity to violations of this assumption, we perform posterior inference on synthetic data with non-independent reporting decisions.

\par For each school $i$ in year $j$, fix the true number of assaults $z_{ij}$ and the reporting probability $p_{ij}$. Given $z_{ij}$ and $p_{ij}$, generate a reported number of assaults $x_{ij}$ under different assumptions:

\begin{itemize}
    \item \textbf{Independent:} For each assault at school $i$ in year $j$, assume the reporting decision is independent with probability $p_{ij}$ of reporting.
    \item $\boldsymbol{\rho}$\textbf{=0.05:} Assume that each assault at school $i$ in year $j$ has marginal probability $p_{ij}$ of being reported, but that the reporting decisions are correlated with correlation coefficient $\rho=0.05$, rather than being independent. Simulate this by sampling $z_{ij}$-many correlated Bernoulli variables, each with probability $p_{ij}$, and taking their sum to get $x_{ij}$.
    \item $\boldsymbol{\rho}$\textbf{=0.10:} Same as above, but with correlation coefficient $\rho=0.10$.
    \item \textbf{Pairwise:} Split the total number of assaults at school $i$ in year $j$ into pairs, and assume that each \textbf{pair} of assaults is reported independently with probability $p_{ij}$. The marginal reporting probability is still $p_{ij}$ for each assault. This creates correlation in the reporting decisions, but unlike the previous scenario, the amount of correlation varies (schools with a large number of assaults will be less affected by this coarsening of the reported numbers, compared to schools where the true number of assaults is low).
\end{itemize}

\begin{figure}[H]
\includegraphics[width=8cm]{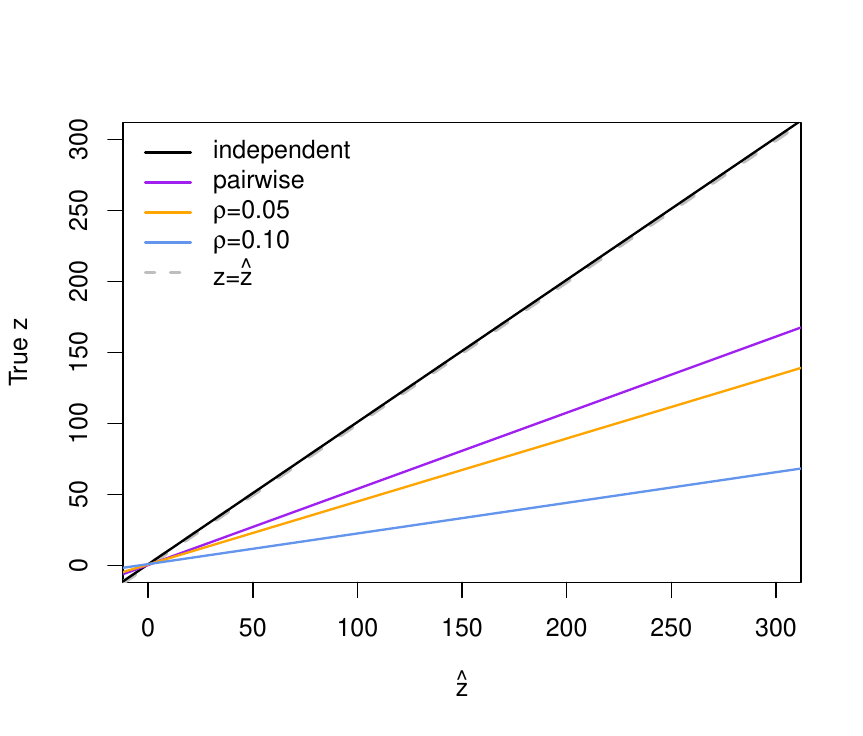}
\caption{When reporting decisions within each school are positively correlated, the model tends to overestimate the true underlying number of assaults.}
\label{fig:correlated_reporting}
\end{figure}

To assess the impact of correlated reporting decisions, we obtain posterior estimates of the true number of assaults for each record in the dataset under each data-generating scenario. For each data-generating scenario, we regress the true number of assaults $z$ on the posterior mean estimate $\hat{z} \overset{\triangle}{=}\mathbb{E}{[z|x]}$. For the dataset simulated with independent reporting decisions, the slope of this regression line is 1.00, indicating that the posterior means of the number of assaults  tend to be proportional to the true counts $z_{ij}$. Figure \ref{fig:correlated_reporting} depicts that the three scenarios with correlated reporting decisions produce slopes less than 1, indicating that the model tends to overestimate the true number of assaults when reporting decisions are positively correlated. This form of model misspecification can significantly degrade the model's ability to recover the true number of assaults.

%%%%%%%%%%%%%%%%%%%%%%%%%%%%%%%%%%%%%%%%
%%%%%%%%%%%%%%%%%%%%%%%%%%%%%%%%%%%%%%%%

\newpage
\section{Sensitivity to Prior Specification}
\label{app:sensitivity}
While the National Crime Victimization Survey reflects a significant effort to measure the true extent of sexual assault victimization in the United States, it is nevertheless true that survey respondents may choose not to disclose their sexual assault victimization to interviewers. Consequently, it is worthwhile to examine how a change in the priors informed by NCVS estimates affects the modeling results presented in Section 5. Below we consider a range of alternative scenarios.

\begin{itemize}
    \item[(a)] Original prior specification presented in Section 4.1-4.2 (Figure 4 and Figure 6).
    \item[(b)] True reporting rates are approximately 10\% lower than NCVS estimates. Prior distributions are adjusted such that the prior median reporting rate is approximately 19.7\%, rather than the 22\% presented in Section 4.2 (Figure 6). Prior distribution on total incidence of assaults is increased correspondingly.
    \item[(c)] True reporting rates are approximately 25\% lower than estimated by NCVS, with corresponding increase in incidence of assaults.
    \item[(d)] True reporting rates are approximately 50\% lower than estimated by NCVS, with corresponding increase in incidence of assaults.
    \item[(e)] True reporting rates are approximately 75\% lower than estimated by NCVS, with corresponding increase in incidence of assaults.
\end{itemize}

\begin{figure}[H]
\begin{center}
    \includegraphics[width=12cm]{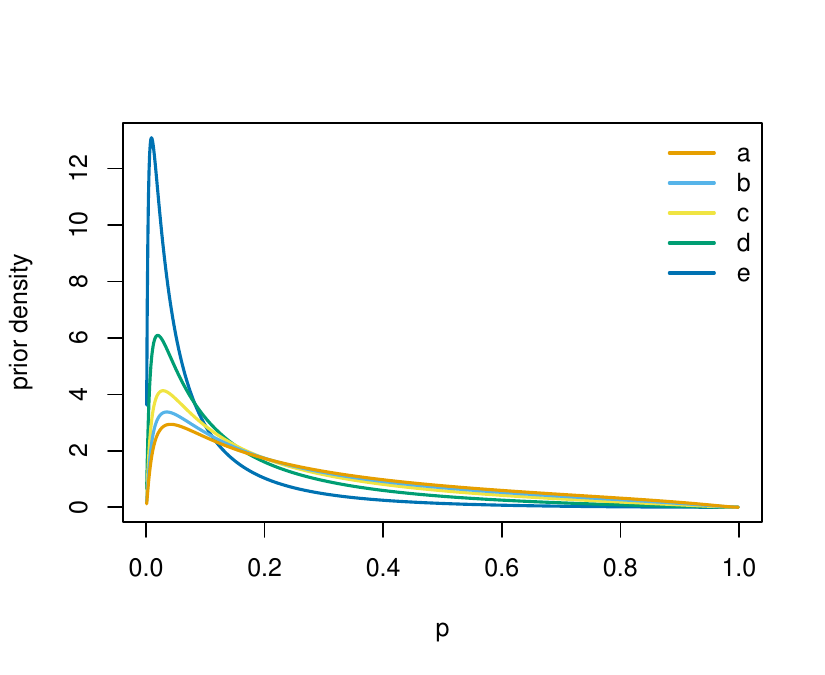} 
    \caption{Prior distributions on the reporting probability $p$ under scenarios (a) - (e) (with model coefficients $\alpha_1$, $\alpha_2$, $\alpha_3$, and $\alpha_4$ set to zero).}
\label{fig:p_prior_alternatives}
\end{center}
\end{figure}

\begin{figure}[H]
\begin{center}
\begin{tabular}{c c}
    \includegraphics[width=6.5cm]{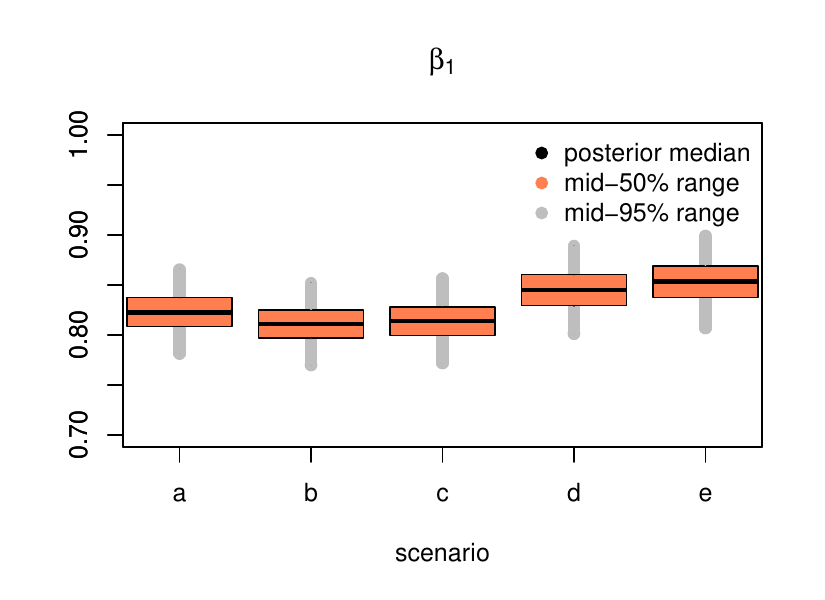} & \includegraphics[width=6.5cm]{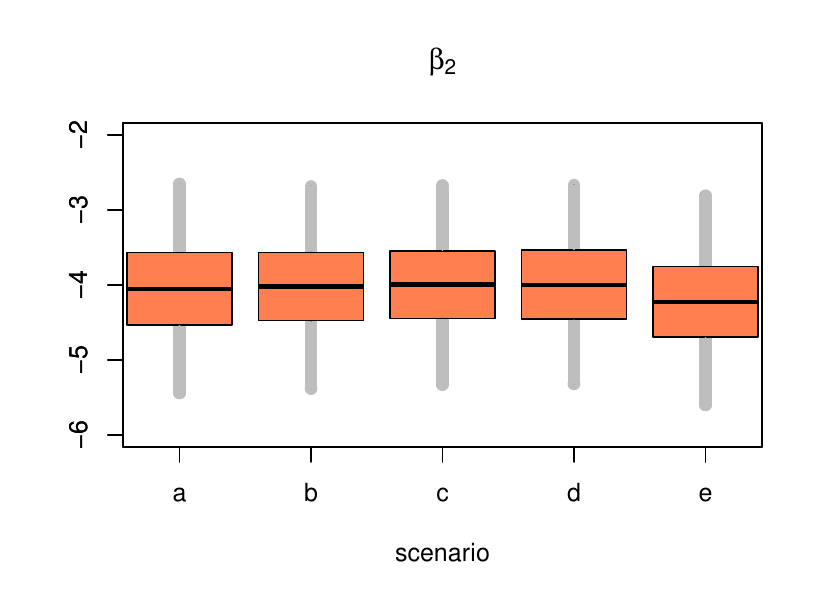} \\
    (i) & (ii)
\end{tabular}
    \caption{Coefficients relating covariates to the total number of assaults do not display much practical difference under moderate changes to the prior distributions envisioned in scenarios a-e. Panel (i) depicts $\beta_{1}$, the coefficient corresponding to student population. Panel (ii) depicts $\beta_{2}$, the coefficient corresponding to gender imbalance of student body.}
\label{fig:beta_boxplots}
\end{center}
\end{figure}

\begin{figure}[H]
\begin{center}
\begin{tabular}{c c}
    \includegraphics[width=6.5cm]{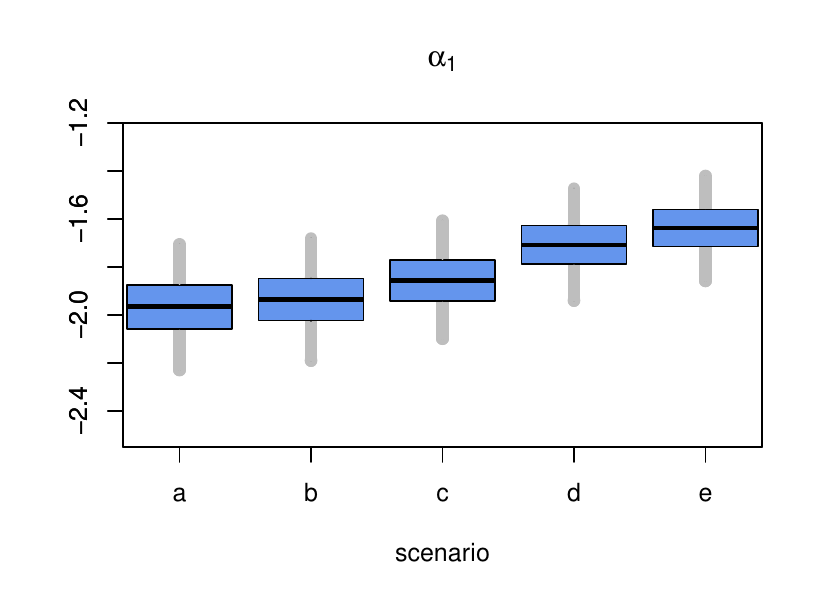} & \includegraphics[width=6.5cm]{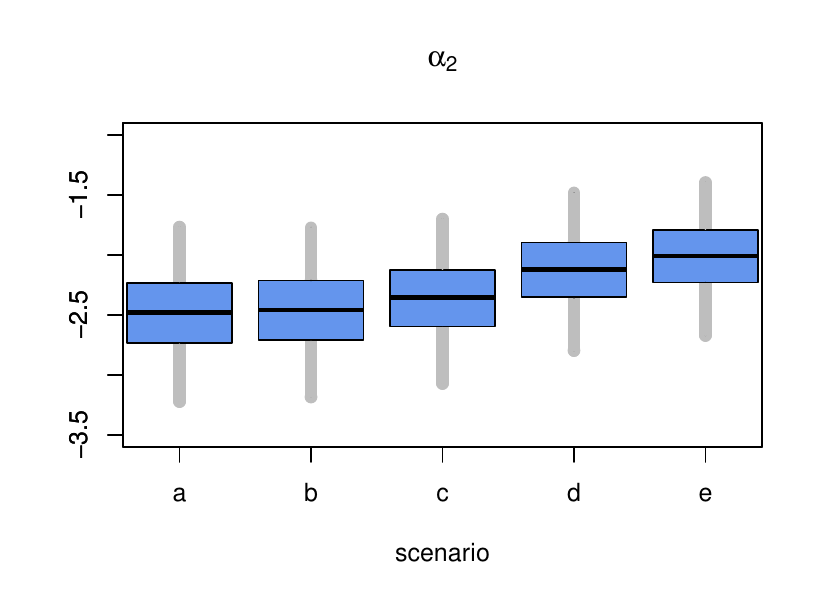} \\
    (i) & (ii)
\end{tabular}
\begin{tabular}{c c}
\includegraphics[width=6.5cm]{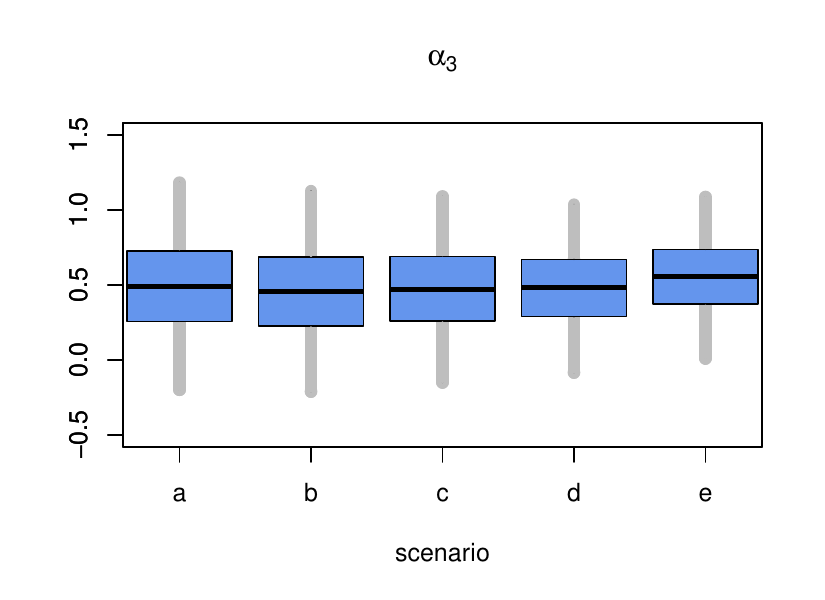} & \includegraphics[width=6.5cm]{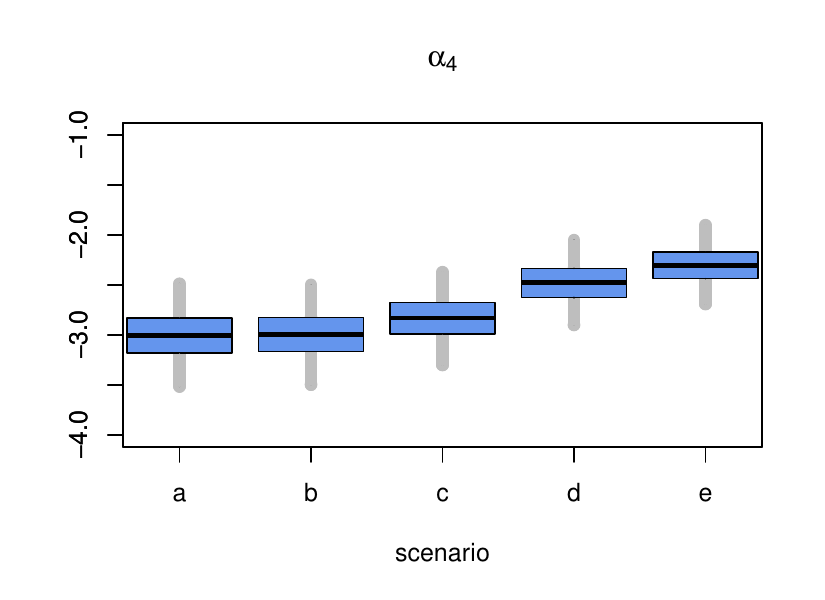}\\
(iii) & (iv)
\end{tabular}
    \caption{Estimates of coefficients relating covariates to the reporting probability remain similar under moderate changes to the prior distributions, and begin to shrink toward zero under more extreme changes to the prior distribution.  Panel (i) depicts $\alpha_{1}$, the coefficient corresponding to whether a school is a junior college. Panel (ii) depicts $\alpha_{2}$, the coefficient corresponding to whether a school is substantially engaged in religious instruction. Panel (iii) depicts $\alpha_{3}$, the coefficient corresponding to the proportion of a school's student body composed of women. Panel (iv) depicts $\alpha_{4}$, the coefficient corresponding to the proportion of a school's student body receiving Pell grants.}
\label{fig:alpha_boxplots}
\end{center}
\end{figure}

\begin{figure}[H]
\begin{center}
    \includegraphics[width=10cm]{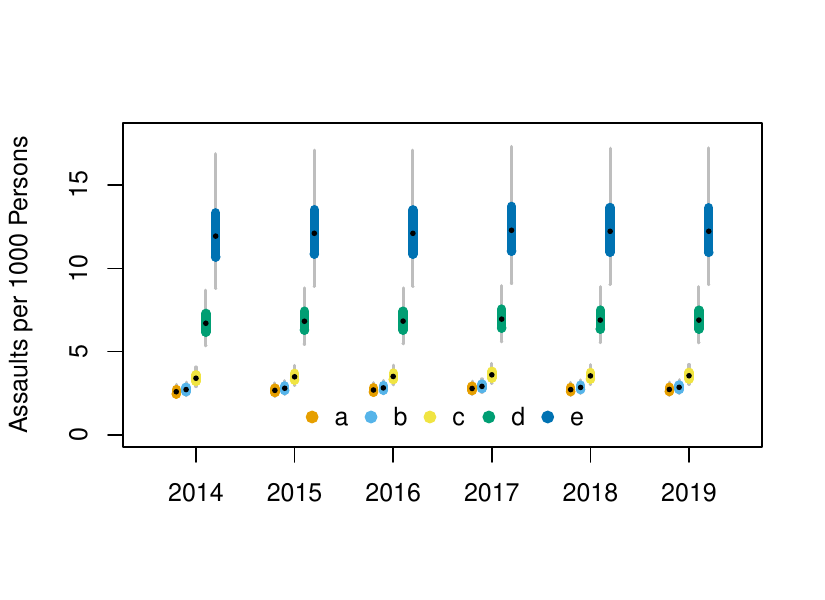} 
    \caption{Prior belief in a higher incidence of sexual assault produces higher posterior estimates of the total number of assaults occurring.}
    \label{fig:sensitivity_incidence}
\end{center}
\end{figure}

\begin{figure}[H]
\begin{center}
    \includegraphics[width=10cm]{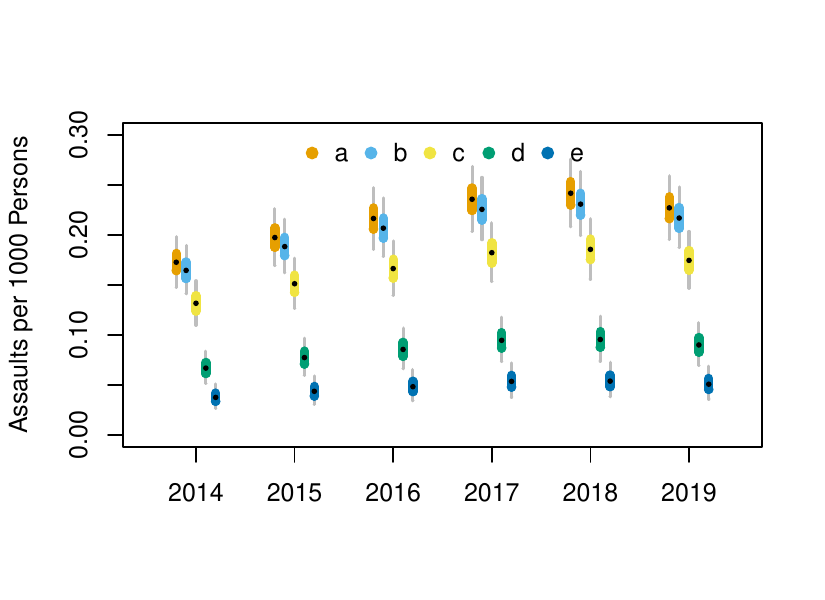} 
    \caption{Prior belief in a lower probability of reporting sexual assault produces lower posterior estimates of the reporting rate.}
    \label{fig:sensitivity_reprate}
\end{center}
\end{figure}

\begin{figure}[H]
\begin{center}
\begin{tabular}{c c}
    \includegraphics[width=6.5cm]{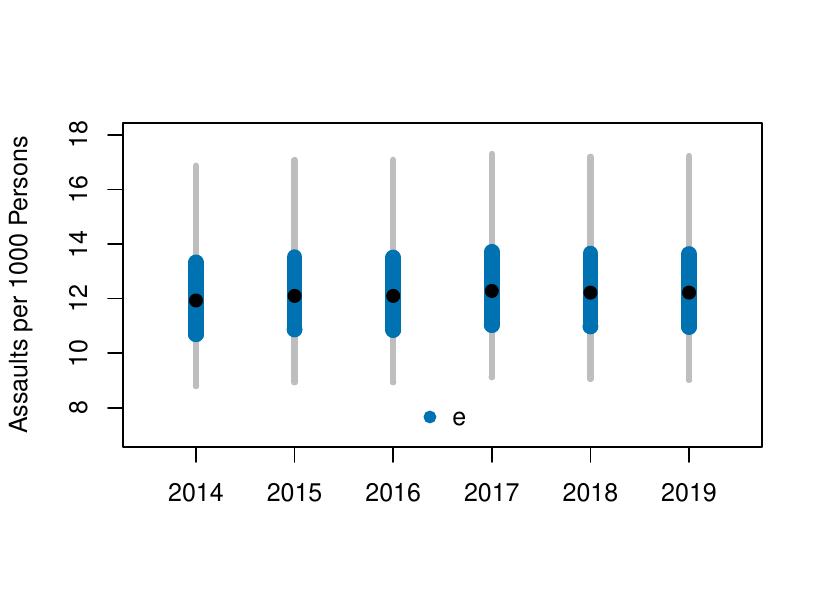} & \includegraphics[width=6.5cm]{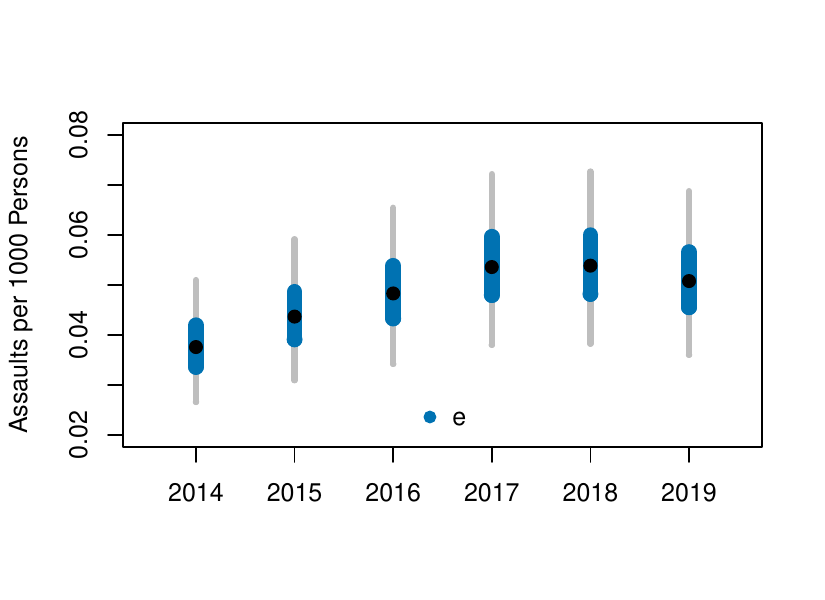} \\
    (i) & (ii)
\end{tabular}
    \caption{Under scenario (e), posterior estimates of the incidence rate do not exhibit a clear increasing or decreasing trend, while posterior estimates of the reporting rate are more suggestive of an increase over 2014-2018. Panel (i) depicts the posterior distribution of the incidence rate; panel (ii) depicts the posterior distribution of the reporting rate.}
\label{fig:sensitivity_aggregate_zoom}
\end{center}
\end{figure}

In line with the discussion in Section 4.5, we find that inferences about the role of covariates are relatively more stable across scenarios (as depicted in Figures \ref{fig:beta_boxplots} and \ref{fig:alpha_boxplots}), while inferences about the true incidence and reporting rate exhibit more sensitivity. As shown in Figure \ref{fig:sensitivity_incidence}, prior belief in a higher incidence of sexual assault produces higher posterior estimates of the total number of assaults occurring. Likewise, Figure \ref{fig:sensitivity_reprate} demonstrates that prior belief in a lower probability of reporting sexual assault produces lower posterior estimates of the reporting rate. However, zooming in on scenario (e) (which represents the most extreme change to the priors) in Figure \ref{fig:sensitivity_aggregate_zoom}, results still suggest that an increase in reporting rates is a more plausible explanation for the increase in the increase in reported assaults over 2014-2018, as opposed to an increase in the true incidence of assault. Note that material variability in NCVS survey respondents' truthfulness over time could call this result into question as well; interested readers may find discussion of a similar issue in the context of marijuana consumption surveys in \citet{cuellar2018trends}.

Given the sensitivity to prior information, we positively note that in 2011 the Bureau of Justice Statistics convened a panel of statisticians, sociologists, and criminal justice experts to review its methodology and recommend optimal practices for measuring sexual assault \citep{kruttschnitt2014estimating}. The panel's recommendations led to the development and pilot testing of new survey approaches to measuring sexual assault within the NCVS context \citep{cantor2021methodological}. These findings fed into a broader multi-year redesign of the overall survey. After further field testing to assess the impact of proposed changes, the redesigned survey is being phased in beginning in 2024 (further detail is available in \citet{truman2023update}).

\end{appendix}

\end{document}